\documentclass[acmsmall,manuscript]{acmart} 
\renewcommand\footnotetextcopyrightpermission[1]{} 
\usepackage{booktabs}
\usepackage{graphicx}
\usepackage{listings}
\usepackage{xcolor}
\usepackage{array}
\usepackage{ragged2e}
\usepackage{hhline}
\usepackage{colortbl}
\usepackage{todonotes}
\usepackage{svg}
\usepackage{amsmath}
\usepackage{enumitem}
\usepackage{setspace}

\newenvironment{fignote}{\begin{quote}\footnotesize}{\end{quote}}
\AtBeginDocument{%
  \providecommand\BibTeX{{%
    \normalfont B\kern-0.5em{\scshape i\kern-0.25em b}\kern-0.8em\TeX}}}

\newcommand{\pseudosection}[1]{\textbf{#1}}

\begin{document}

\title[The Who in XAI] {The \textit{Who} in XAI: How AI Background Shapes Perceptions of AI Explanations}

\author{Upol Ehsan}
\affiliation{%
  \institution{Georgia Institute of Technology}
  \city{Atlanta}
  \state{GA}
  \country{USA}}

\author{Samir Passi}
\affiliation{%
  \institution{Microsoft}
  \state{}
  \country{USA}}
 \email{}

\author{Q. Vera Liao}
\affiliation{%
  \institution{Microsoft Research}
  \city{Montreal}
  \state{}
  \country{Canada}}

 \author{Larry Chan}
\affiliation{%
  \institution{Georgia Institute of Technology}
  \city{Atlanta}
  \state{GA}
  \country{USA}}

 \author{I-Hsiang Lee}
\affiliation{%
  \institution{Georgia Institute of Technology}
  \city{Atlanta}
  \state{GA}
  \country{USA}}

\author{Michael Muller}
\affiliation{%
  \institution{IBM Research AI}
  \city{Yorktown Heights}
  \state{NY}
  \country{USA}}

 \author{Mark O. Riedl}
\affiliation{%
  \institution{Georgia Institute of Technology}
  \city{Atlanta}
  \state{GA}
  \country{USA}}


\renewcommand{\shortauthors}{Ehsan, Passi, Liao, Chan, Lee, Muller, and Riedl}

\begin{abstract}
Explainability of AI systems is critical for users to take informed actions. Understanding \textit{who} opens the black-box of AI is just as important as opening it. We conduct a mixed-methods study of how two different groups—people with and without AI background—perceive different types of AI explanations. Quantitatively, we share user perceptions along five dimensions. Qualitatively, we describe how AI background can influence interpretations, elucidating the differences through lenses of appropriation and cognitive heuristics. We find that (1) both groups showed unwarranted faith in numbers for different reasons and (2) each group found value in different explanations beyond their intended design. Carrying critical implications for the field of XAI, our findings showcase how AI generated explanations can have negative consequences despite best intentions and how that could lead to harmful manipulation of trust. We propose design interventions to mitigate them.

\end{abstract}

\ccsdesc[500]{Human-centered computing~Empirical studies in HCI}
\ccsdesc[500]{Human-centered computing~User studies}
\ccsdesc[300]{Human-centered computing~Empirical studies in collaborative and social computing}
\ccsdesc[500]{Computing methodologies~Artificial intelligence}

\keywords{Explainable AI, Human-Centered Computing, Data Vision, User Perceptions, Artificial Intelligence, Heuristics, Appropriation, User Characteristics}

\maketitle
\pagestyle{plain} 

\section{Introduction}
As AI-driven systems increasingly power high-stakes decision-making in public domains such as healthcare~\cite{holzinger2017we,katuwal2016machine,che2016interpretable,loftus2020artificial}, finance~\cite{MacKenzie2018, murawski2019mortgage}, law~\cite{tan2018distill,veale2017fairer, berk2013statistical}, and criminal justice~\cite{Rudin2020, Kleinberg2017, hao2019jail}, their explainability is critical for end-users to take informed and accountable actions~\cite{smith2020no}. 
Issues concerning explainability lie at the heart of Explainable AI (XAI), a research area that aims to provide human-understandable justifications for the system's behavior~\cite{ehsan2019automated,adadi2018peeking,guidotti2018survey}.
Explainability is not a new issue within AI~\cite{shortliffe2012computer, moore1988explanation, johnson1994agents}, but the proliferation of Deep Learning and Reinforcement Learning based approaches---models of which are considered hard to interpret, even by experts---has led to remarkable growth in techniques that aim to ``open'' the AI opaque box~\cite{guidotti2018survey}. 

While opening the opaque box is important, {\em who} interacts with the box also matters. Implicit in XAI is the question: “explainable to whom?” ~\cite{ehsan2020human}. 
The \textit{who} governs the most effective way of describing the \textit{why} behind the decisions.
Getting a situated understanding of how different \textit{who's} with different user characteristics matter in XAI is thus important. 
To give an illustrative example: riders (end-users) of a self-driving car have different user characteristics than its engineers (developers). Riders, many of whom are not AI experts, might not have the AI background that the engineers have and thus have different explainability needs and goals. 

One's AI background is an impactful user characteristic in XAI because there is often a disparity in this characteristic between creators/developers and end-users, which can lead to usability failures and irresponsible design~\cite{norman2013design}, even inequities~\cite{chromik2021think}. 
Many end-users are unlikely to have AI backgrounds comparable to the creators of the technology~\cite{jin2019bridging}. Nonetheless, XAI developers tend to design explanations  \textit{as if} people like them are going to use their systems~\cite{miller2019explanation}. In fact, a majority of current deployments of XAI technologies serve AI engineers instead of end-users~\cite{liao2020questioning,alqaraawi2020evaluating}. This creates a consumer-creator gap, one between design intention and reality---how developers envision the AI explanations to get interpreted and how users actually perceive them. If we want to bridge this gap, we need to understand how user characteristics, such as AI background, impact it.\footnote{By highlighting the creator-consumer gap in XAI, we do not mean to undermine the diversity of stakeholders in the ecosystem. By calling attention to extreme ends, we are highlighting the severity of the gap while fully acknowledging the ecosystem's diversity. See, e.g.: \cite{Miceli2020, gray2019ghost, Muller2019HC, Passi2020}.}

In this paper, we share \textit{how} and \textit{why} one's AI background (or lack thereof) shapes their perceptions of AI explanations. Focusing on two groups, one with and one without an AI background, we found that~(1)~both groups had unwarranted faith in numbers, but exhibited it for different reasons and to differing degrees, with the AI group showing a higher propensity to over-trust numerical representations and potentially be misled by the presence of it. 
(2)~Each group found explanatory values in different explanations that went beyond the usage we designed them for.
These insights have potential negative implications like susceptibility to harmful manipulation of user trust. 

We found these insights through a mixed-methods study where we probed for user perceptions of \textit{three} types of AI-generated explanations: 
(1)~natural language with justification (explaining the “why” behind the action), 
(2)~natural language without justification (describing “what” the action was), and 
(3)~numbers that determine the agent’s actions (akin to ``transparent'' AI). 
We measure perceptions along five dimensions: \textit{confidence, intelligence, understandability, second chance,} and \textit{friendliness}, which are grounded in related work around HCI, HRI, and XAI~\cite{davis1989user,venkatesh2003user,ehsan2019automated, binns2018s, das2020leveraging} and quantitatively share within- and between-group differences. 
Through qualitative analysis, we examined \textit{how} AI background shaped each group's interpretation of explanations and highlight \textit{why} their perceptual differences might exist.

We elucidate the \textit{why} behind the group differences using the conceptual lenses of \textit{heuristics} (mental shortcuts)~\cite{kahneman2011thinking,sherman1984cognitive} and \textit{appropriation} (users' repurposing of a design)~\cite{dix2007designing, muller2016collaborative, salovaara2008inventing}. In light of the findings, we share concrete design implications around mitigating the risks of overreliance on numbers which can potentially lead to negative consequences such as over-trust on XAI systems. 
We share broader lessons around how our insights can help re-imagine AI education and mitigate potential harmful manipulation with explanations. By bringing conscious awareness of the group differences to the human-centered design of XAI systems, we address the AI creator-consumer gap by making the following contributions:
\begin{itemize}
    \item We quantify user preferences (\textit{what)} of three types of AI explanations along five dimensions of user perceptions.
    \item We qualitatively situate \textit{how} one's AI background (or lack thereof) influences the perception of the explanations.
    \item We elucidate \textit{why} the group differences or similarities might exist and interpret them through the conceptual lenses of heuristics and appropriation.
    \item Using our findings, we identify potentially negative consequences (like harmful manipulation of user perceptions and over-trust in XAI systems) and propose mitigation strategies. 
\end{itemize}

\section{Background}\label{background_section}
In this section, we review related work in the field of XAI salient to the paper, highlight the need to attend to XAI's sociotechnical dimensions and human-centered perspectives, and discuss HCI work studying how user background shapes users' perception of and needs for technology that motivated our work.

\subsection{Explainable AI}
While the origin of ``explainable AI'' can be traced back to expert systems in the 1980s~\cite{swartout1983xplain}, the field of XAI has been undergoing a resurgence due to the proliferation of complex Deep Learning models. Although there is a current lack of consensus on the meaning of explainability and related terms such as interpretability~\cite{rosenfeld2019explainability,arrieta2020explainable}, XAI work shares a common goal of making the AI systems' decisions or behaviors understandable by people~\cite{ehsan2019automated,adadi2018peeking}
Among other dimensions to map the landscape of technical XAI approaches, the field differentiates between methods to build directly interpretable models and methods to generate explanations for opaque-box models~\cite{lipton2001good,ribeiro2016should, yosinski2015understanding,gilpin2018explaining} (for a detailed overview see recent survey papers~\cite{guidotti2018survey,arrieta2020explainable,adadi2018peeking}). While simpler models such as linear regression and decision-tree are typically considered directly interpretable but low-performing, recent work (e.g.~\cite{dash2018boolean,wei2019generalized}) focuses on developing new algorithms that produce ``clear-box'' models allowing ``under the hood'' inspection without sacrificing performance.

In contrast, \textit{explanation generation} methods--used in this paper--aim to explain models that are not directly human-understandable (e.g., deep neural networks). They are often post-hoc techniques~\cite{lipton2018mythos, miller2019explanation, ribeiro2016should, yosinski2015understanding, ehsan2019automated} that could be applied after model building. Typically, these methods rely on distilling a simpler model from the input and output~\cite{ribeiro2016should,lundberg2017unified} or meta-knowledge about the model~\cite{ehsan2019automated} to generate explanations that approximate the model's behavior. While there is, by design, a loss of scrutability, these methods allow the flexibility to make any model explainable, and thus have become popular and been applied to transforming simulation logs to explanations\cite{van2004explainable}, intelligent tutoring systems \cite{core_building_2006}, transforming AI plans into natural language \cite{vanlent:aiide2005}, and translating multi-agent communication policies into natural language \cite{andreas2017translating}. By privileging accessible understanding over revealing ``under the hood'' model mechanisms, explanation generation methods can be geared toward non-AI experts.
In this paper, we focus on a specific explanation generation technique called \textit{rationale generation}~\cite{ehsan2019automated}---a process of producing a natural language explanation for agent behavior as if a human had performed the behavior and verbalized their inner monologue. 
While explanations can be in any modality, rationales are natural language-based, making it especially accessible for non-AI experts \cite{ehsan2019automated}.

\subsection{Towards Human-Centered XAI}\label{background_hcxai}
There has been a growing recognition that XAI systems are often developed without an understanding of the recipients' needs and characteristics~\cite{ehsan2020human,miller2019explanation}. 
For instance, many of the XAI techniques created to support explainability needs during model development~\cite{ribeiro2016should,selvaraju2017grad} may break down when it comes to serving end-users with different needs~\cite{liao2020questioning}. 
It is imperative to follow human-centered approaches to understand the “personal, social, and cultural aspects” \cite{jaimes2007guest} of the recipients of AI explanations, especially since a monolithic view of the \textit{who} may inadvertently risk dehumanization \cite{chancellor2019human,kling1998human}. Given deployment in high-stakes settings, AI systems designed without attending to the needs and values of different stakeholders may also risk marginalizing certain groups or exacerbating existing inequities~\cite{barocas2016big, fairness2019}.

The need for human-centered approaches in XAI has inspired increasing efforts among HCI and CSCW researchers, following the community's long-standing tradition of designing and studying explainable computing systems~\cite{Lim2009,kulesza2013too,Eslami2015,Rader2015,eiband2021support}. 
Studies empirically evaluating XAI techniques in specific use contexts reveal the divergent needs and preferences of users~\cite{Kaur2020,dodge2019explaining,cheng2019explaining,cai2019effects,hohman2019gamut,alqaraawi2020evaluating}.
For example, while data scientists might need multiple XAI tools for a comprehensive understanding~\cite{hohman2019gamut}, simple explanations are often sufficient for AI-novices~\cite{cheng2019explaining}.
User characteristics such as cognitive load disposition~\cite{ghai2021explainable} and general trust in AI~\cite{dodge2019explaining} could moderate how users perceive AI explanations. 
Some studies further reveal potential drawbacks of AI explanations--- how explanations could impose undesired cognitive burden~\cite{abdul2020cogam}, create a false sense of security and over-trust~\cite{Kaur2020,ghai2021explainable}, and how even placebic explanations (devoid of justificatory content) can engender trust in AI systems~\cite{Eiband2019}.

Researchers have begun to examine people's cognitive process of interpreting AI explanations, which could help us understand atypical and misaligned user receptions of XAI. 
Recent work highlights the dual-process of cognition when people process AI explanations ~\cite{buccinca2020proxy}. 
The dual-process theory~\cite{wason1974dual,kahneman2011thinking,petty1986elaboration} posits that people's cognitive processes follow two systems: System 1 processes stimuli in a fast and automatic manner, whereas System 2 engages in deliberative and analytical thinking. System 1 often relies on heuristics (rules-of-thumb or mental shortcuts) that can be developed through past experiences. These heuristics, if applied inappropriately, should be considered cognitive biases~\cite{kahneman2011thinking}. 
In XAI, there is often an assumption that people mostly engage in analytic System~2 thinking whereas there is growing evidence that people mostly engage in System~1 thinking~\cite{buccinca2020proxy}.
Negative consequences of cognitive biases such as over-trust in XAI could be attributed to a System~1 heuristic of associating explanations with AI competence~\cite{nourani2021anchoring,chromik2021think}. One way to mitigate these biases would be to use \textit{Cognitive Forcing Functions}~(CFFs)--interventions that disrupt heuristic reasoning and promote System 2 analytical thinking~\citep{croskerry2003cognitive}.

Human-centered XAI calls for pluralistic explanation design---for the \textit{who}, not just the \textit{what}, of XAI~\cite{ehsan2020human}. Recent XAI work has begun to differentiate major categories of XAI consumers, from builders to regulatory bodies~\cite{arrieta2020explainable,tomsett2018interpretable,mohseni2020goals,dhanorkar2021needs}. 
While this work is informative, there is still a dearth of empirical insights and understanding of the differences between these XAI users, and actionable design guidelines to support them. 

A generative approach towards pluralistic design of XAI, we believe, is to develop a systematic understanding of \textit{how} and \textit{why} users with different characteristics (e.g., with or without AI backgrounds) form different perceptions (\textit{what}) of XAI systems. Such insights can refine our understanding of \textit{who} the humans are in XAI. 
To our knowledge, this has not yet been systematically explored in the specific context of XAI. Thus, our work adds to the discourse by adding empirical insights through a systematic exploration of two different \textit{who's} in XAI. Our work is also motivated by broader research studying individual differences that impact human-AI interaction, as reviewed below.

\subsection{Individual differences in human-AI interaction}
\label{sec:individual-differences}
There is a rich history of transforming insights into how individual differences impact technology perception into the design of personalized, accessible, and inclusive technologies. For example, there is a large body of work on how various types of epistemic background, including computer literacy~\cite{disessa2001changing}, digital literacy~\cite{bawden2008origins}, and numeracy~\cite{jensen2010utilization}, impact user competency to use computing systems and ways to mitigate the competency gaps. 

Recent work has paid attention to how user characteristics impact human-AI interaction.
For example, studies on human-robot~\cite{lee2010receptionist,das2021explainable,stange2020effects} and human-agent interaction~\cite{liao2016can,liao2018all} found that user's schema, whether an agent is seen as a utilitarian tool or a social entity, leads to noticeable differences in interactions with and evaluations of the agent.
Research around AI fairness has identified various mediating factors such as users' education level, fairness criteria, and general trust in ML~\cite{dodge2019explaining,wang2020factors}. 
In short, individual differences and user characteristics shape perceptions of AI systems and should be appropriately understood and carefully accommodated when designing them. 

A commonly studied user characteristic in Human-AI interaction is users' knowledge about AI, often operationalized as AI programming experience~\cite{myers2019impact}, building ML models~\cite{ghai2021explainable}, or type of profession (e.g., data scientist)~\cite{hohman2019gamut}. 
Recently Long and Magerko provided a concrete definition of AI literacy~\cite{Long2020} with a set of core competencies to understand and use AI, and proposed design guidelines to mitigate the competency gaps. 
Recent work also examines the implications of background in AI as a determining factor of one's \textit{role} in an AI ecosystem. 
Motivated to ``problematize the asymmetric relationship between technical experts and users'', Cheon and Su~\cite{cheon2017configuring} highlighted the misalignment between the creators' (roboticists') vision of how consumers (end-users) interpret and use the product--a salient theme in this paper as well. 
McDonald and Pan~\cite{mcdonald2020intersectional} examined how CS students (likely to become AI developers) viewed ethical problems in AI, found substantial limitations, and called for a closer integration of ethics education with technical training. 

Our understanding of how people's AI backgrounds might impact their perceptions of AI explanations also draws from a long line of social science research on the relations between sensemaking \cite{Weick1995} and professional knowledge \cite{Goodwin1994}. Passi and Jackson \cite{Passi2017}, for instance, analyzed academic learning practices in the field of data science to show how students start ``seeing'' the world differently once they learn to work with algorithms and numbers. Through ethnographic fieldwork, they highlight how having a computational background enables students to gain actionable forms of ``data vision''---ways of seeing that allow them to approach and analyze the world as data.

A person's AI background (or a lack thereof)---a focal point of this paper---in fact, directly impacts user perceptions. In a recent ethnographic study~\cite{Passi2018}, researchers found that data scientists and business analysts perceived an AI system's accuracy score differently: business analysts saw the score as a measure of overall performance (good vs. bad), while data scientists perceived more granular insights into types of errors (false positives vs. false negatives). As people learn specific ways of \textit{doing}, it also changes their own ways of \textit{knowing}---in fact, as we argue in this paper, people’s AI background impacts their perception of \textit{what} it means to explain something and \textit{how}.

Thus, the \textit{who} questions are important, because people tend to interpret technologies differently, leading to different usages of those technologies \cite{anacleto2013adoption, bowers2012logic, dix2007designing, muller2016collaborative, salovaara2008inventing}.
The process of differential interpretation, followed by different usage, has been called \textit{appropriation}--e.g., usage patterns that go beyond the original designers' expectations~\cite{skrischkowsky2018thesis, tscheligi2014potentials}.
Dix listed six principles of user appropriation of technologies, including support for interpretive flexibility~\cite{dix2007designing}. In this paper, we go beyond the insight that different people need different explanations. Building on the discourse of user characteristics in XAI~\cite{suresh2019framework, nourani_importance_2022}, we extend the literature through an in-depth exploration of how and why interpretation and use of explanations may differ between people with or without AI background.

\section{Study Design and Methods} 
We begin by sharing the research questions (RQs) followed by how we operationalize key aspects of the research design. 

\begin{itemize}
    \item RQ1: Quantitatively, how do people with different AI backgrounds perceive different explanations given by AI agents?
    \item RQ2: Qualitatively, how and why do differences in AI background result in different or similar perceptions of explanations? 
\end{itemize}

We address these RQs by conducting a within-subjects experiment in which two groups of participants, with or without an AI background, see three versions of AI agents (depicted as robots in our study,~Fig.~\ref{fig:3robots}) with different types of explanation. Using this 2x3 factorial design, we quantitatively measure the perceptions of the AI explanations and compare the differences between the two participant groups to address RQ1~(Section~\ref{sec:quant_analysis}). Participants rank their preferences and justify their choices through open-ended text responses, which we qualitatively analyze to understand the underlying differences between the two groups to address RQ2~(Section~\ref{sec:qual_findings}). Below, we unpack how we operationalize three things: (1)~explanation types, (2)~user perceptions, and (3)~backgrounds in AI.

\subsection{Explanation Generation Method and Types} \label{sec:explanation_gen_description}
\begin{figure*}[tbh]
    \caption{The three robots navigating the task environment and explaining their actions. From left to right, the robots and their colors are: Rationale-Generation (in blue), Action-Declaring (in purple), and Numerical-Reasoning (in green).
    In the screenshot, each robot is taking the same action, but they are explaining it differently.
    The explanation text accompanying each robot is taken verbatim from the videos participants watched. To improve legibility, the text has been remastered to a higher resolution.}
    \centering
    \includegraphics[width=11cm]{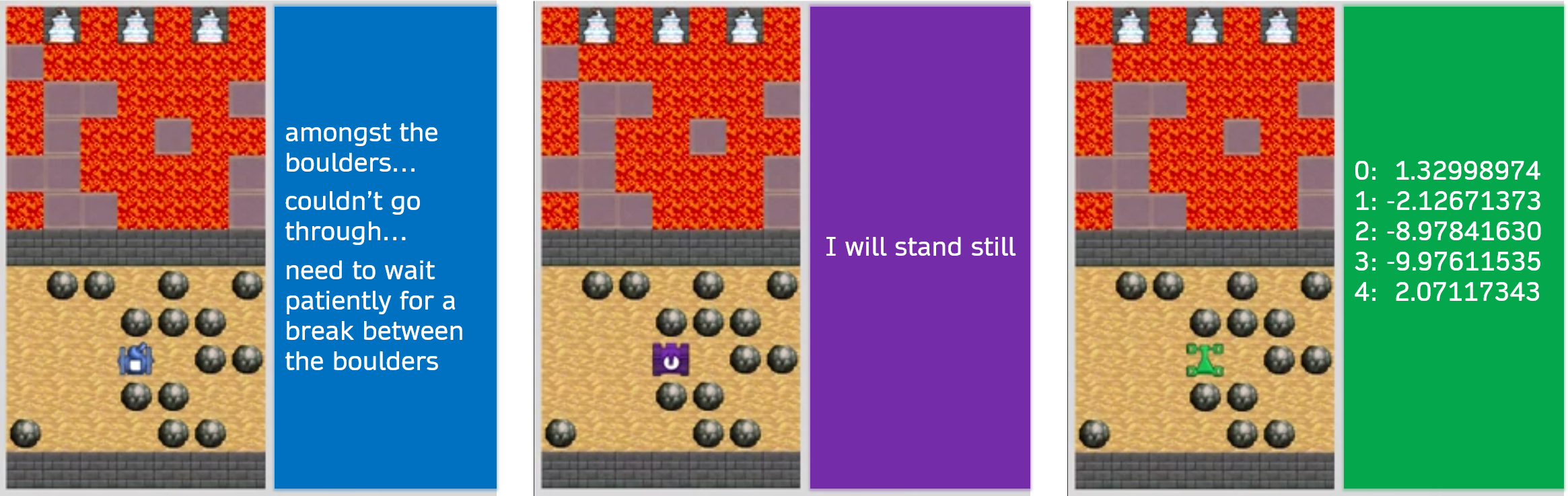}
            \label{fig:3robots}
            \Description[Screenshots of the three robots in the environment.]{Screenshots of the three robots in the environment. The left shows a robot }
    \Description[Screenshots of the three robots operating in the virtual environment.]{Three screenshots, each showing a purple, blue, or green rover navigating through a field of boulders toward a field of lava with floating platforms. The configuration of boulders is the same for each robot. The left screenshot shows the Rationale Generating robot along side text, ``amongst the boulders... couldn't go through... need to wait patiently for a break between the boulders. The middle screenshot shows the Action Declaring robot alongside text, ``I will stand still.'' The right screenshot shows the Numerical Rationale robot alongside text 
    ``0:1.32998974\\
    1: -2.12671373\\
    2: -897841630\\
    3: -9.97611535\\
    4: 2.07117343''}
\end{figure*}

We begin with the \textit{task environment} to situate the design of the AI agents. In the user study (task details in Section \ref{sec:task_details}), participants watched 3 robots (AI agents) carry out an identical sequence of actions which differed only in the way that the AI agent "thinks out loud” about its actions. The robots need to navigate through a sequential decision-making environment---a field of rolling boulders and a river of flowing lava--to retrieve essential food supplies for trapped space explorers~(Fig.~\ref{fig:3robots}). The robots thus need to observe a dynamic environment and think ahead in order to complete an objective. We chose a sequential environment because the explainability of sequential tasks is under-explored, while prior XAI work has explored non-sequential tasks (e.g., classification, captioning, etc.~ \cite{wang2017residual,xu2015show,you2016image}). 
To solve the navigation problem, the agent used a Reinforcement Learning (RL) algorithm called {\em tabular $Q$-learning}~\cite{watkins1992q}. Reinforcement Learning is both a promising technology for autonomous AI systems but also challenging from an explanation perspective. Tabular $Q$-learning agents attempt to learn the utility (called a $Q$-value for ``quality'' of the action) of different actions in different situations. Once learning is complete, the agent makes decisions by picking the action with the highest $Q$-value. In our case, the fully trained agent solved the environment, generating an action trace that contains the sequence of steps needed to reach the goal (food supplies) without failure. 
Recall that our study has 3 robots (AI agents) with different types of explanation. In order to standardize their actions across experimental conditions, \textit{all robots use the same trace}. This means that the decision-making mechanism underlying each robot is the same RL-based one. They \textit{only differ} in how they explain their actions. 

\vspace{-2mm}

\paragraph{Explanation generation:} The three different types of explanation are the within-subject variable in our study comparing perception differences between the AI and non-AI groups. 
Based on a review of related work, we chose three types of explanation that vary in their justification quality and representation modality (e.g., textual, numerical).
With these considerations in mind, we set up the robots to express themselves in three ways: (1)~natural language with justification (explaining the “why” behind the action; Fig~\ref{fig:3robots}, left), (2)~natural language without justification (describing “what” the action was, Fig~\ref{fig:3robots}, center), and (3)~numbers that determine the agent’s actions (akin to ``transparent'' AI, in this case showing $Q$-values, Fig~\ref{fig:3robots}, right). 
We share the explanation mechanisms and the attributes of the three robots below.

\begin{itemize} 
    \item The \textit{Rationale-Generating (RG)} robot (Robot A in the study): this robot “thinks out loud” in natural language rationales explaining the “why” behind the action (\#1 above). Our generation approach is similar to prior work in XAI and HRI~ \cite{das2020leveraging,ehsan2017rationalization} where they use a neural machine translation (NMT) ~\cite{luong2015effective} approach to produce plausible rationales to explain sequential behavior. We extend and adapt it to fit our sequential environment depicting a space mission. The RG robot's expressions aim to provide a \textit{functional} understanding~\cite{lombrozo2014explanation, lombrozo2019mechanistic} of the actions by appealing to functions or goals of the agent (e.g., Fig~\ref{fig:3robots}, left). The RG robot has language and justification.  
    \item The \textit{Action-Declaring (AD)} robot (Robot B in the study): this robot “thinks out loud” by stating its action in natural language without any justification. It's a neutral option that simply states “what” the action is (\#2 above). For instance, it states “I will move right” as it moves right. These outputs are generated from pre-fixed expressions that are triggered based on the agent’s action. 
    \item The \textit{Numerical-Reasoning (NR)} robot (Robot C in the study): this robot “thinks out loud” by simply outputting the numerical Q-values for the current state with no language component (\#3 above). It is akin to a “transparent AI” where we directly look inside the opaque RL-box by observing its Q-values. Q-values can provide some transparency into the agent's beliefs about each action's relative utility (``quality''). However, Q-values themselves do not contain information on ``why'' one action has a higher utility than another. Note that we do not indicate that the numbers are $Q$-values in our study nor indicate which value is associated with which action. 
\end{itemize}

While all robots ``think out loud'', only the RG robot is designed to have any justificatory quality---it is designed to provide a functional understanding~\cite{lombrozo2014explanation} of the ``why'' behind a robot's action. The other two conditions provide baselines that should be considered as lacking justificatory qualities by design. The AD robot merely states ``what'' the action was (a neutral option). While NR could, in theory, provide a mechanistic understanding~\citep{lombrozo2019mechanistic} of the ''how'', the unlabelled numerical format should make its meaning difficult to access, if not impossible. While AD and NR do not have explanatory qualities by design, we do not know if or how participants will interpret them. Through the experiment, we are interested in whether and how these explanations invoke different perceptions in the two groups.

\subsection{Grounding the Dimensions of User Perceptions} \label{sec:user_perceptions}

We now motivate \& ground the dimensions of perceptions. To scope dimensions (measures) of perceptions appropriate for our use case, we engaged in an iterative filtering process---this process included (1) a systematic review of related work around trust, acceptance, and engagement of autonomous or AI systems followed by (2) informal interviews with six experts spanning HCI, AI, and HRI. 
The aforementioned process informed the adaptation of the following dimensions from classical technology acceptance models and emerging work in HRI and XAI literature ~\cite{davis1989user,venkatesh2003user,ehsan2019automated, binns2018s, das2020leveraging}. 

One of the core goals of XAI is to make AI systems more understandable. The improved understandability (or lack thereof) can impact one's trust or confidence in the system. In line with these facets, we adapted \textit{understandability}  and \textit{confidence} from TAM \& UTAUT~\cite{davis1989user,venkatesh2003user}. 
Prior work in HRI and AI shows that tolerance to failure~\cite{desai2013impact} and perceived capability of the AI system ~\cite{koda1996agents} are impacted by how one perceives how intelligent the agent is; thus we added \textit{intelligence} to our list of user perceptions. 
Recent work in autonomous system acceptance~\cite{ sohn2020technology} shows that sociability factors are core markers of user adoption~\cite{lin2012integration}. We adopted the dimension of \textit{friendliness} or how friendly an agent appears because of its impact on relationship development~\cite{huang2014modeling, tickle1990nature} and partnership~\cite{cafaro2012first, bickmore2004towards}, which is essential for human-AI collaboration~\cite{nass1996can, nass1994machines, nass2000machines}.
Emerging work in XAI and HRI~\cite{lee2010gracefully, mirnig2017err,short2010no, kwon2018expressing} suggests that how an AI agent communicates failure governs future collaboration relationships with humans. Thus, we add the notion of a \textit{second chance} to understand how past failures impact future collaboration chances for XAI agents.  
In our study, participants ranked the robots along these \textit{five} perception dimensions and justified their choices using open-ended text responses:
\begin{enumerate}
    \item \textit{Understandability:} Based on their explanations, I found each robot's explanation of its actions understandable in the following order
    \item \textit{Confidence:} Based on their explanations, I would rank my confidence in each robot's ability to do its task in the following order
    \item \textit{Intelligence:} Based on their explanations, I would rank each robot's intelligence in the following order
    \item \textit{Friendliness:} Based on their explanations, I would rank the friendliness of each robot in the following order
    \item \textit{Second chance:} Each robot failed.  Based on their explanations, I'd rank my willingness to give another chance to each robot in the following order

\end{enumerate}

\subsection{Participants: Operationalizing AI vs. non-AI Backgrounds } \label{participants_section}

We now address how we operationalized the user background in our study.
We acknowledge that AI background can be operationalized in different ways. 
As a \textit{formative first step}, we use a ``high contrast'' approach where there is a stark difference between the groups.
That is why, we focus on people \textit{with} and \textit{without} an AI background. 
This high-contrast approach provides a baseline understanding and creates the foundation for future granular operationalizations (e.g., years of AI experience). 
Below we share the motivation and operationalization of each group formation.

\subsubsection{The AI Background Group}
\hfill\\
For the AI group, we recruited participants who are students enrolled in CS programs and taking AI courses. Granted there are other ways to operationalize this group (e.g., recruiting AI practitioners), as a first step, we chose students because it allows us to explore how their AI coursework impacts the way they make sense of explanations from AI systems. Since professionalization starts with academic training~\cite{Passi2017}, investigating the roots of one's AI background and the impact on how students make sense of AI systems is important. In fact, as we show in this paper, \textit{especially} during their learning phase, AI students adopt reasoning artifacts that impact their perceptions of explanations from AI systems in very specific ways, which has core implications for the future of XAI design (a point we elaborate in Section \ref{sec:broader_lessons}). Granted that not every AI student will go on to build AI systems but with the proliferation of AI systems in the workplace, a majority of these students could become stakeholders residing on the creation or development end of the technology spectrum---as potential developers, designers, and managers of AI-based systems. As potential creators of AI systems, their perceptions matter in bridging the creator-consumer gap in XAI. 

\subsubsection{The Non-AI Background Group}
\hfill\\
For the non-AI group, we recruited participants from Amazon Mechanical Turk (AMT).
Carefully screened AMT participants have been shown to be representative of consumer research~\cite{bentley2017comparing, kittur2008crowdsourcing, goodman2017crowdsourcing}, which facilitates our goals of comparing potential creators (the AI group) with potential end-users (the non-AI group). 
We acknowledge that consumers too can and do have significant AI backgrounds, which is why we systematically screen out people in the non-AI group with any level of AI knowledge (using the screening process outlined in \ref{sec:screening_and_recruit}).
Non-AI students, albeit an intuitive comparison group, would be a subset of the larger consumer base making them a good candidate for future granular investigations.
Multi-disciplinary research has also shown how AMT participants can be reasonable alternatives to a university participant pool in terms of data integrity~\cite{sprouse2011validation, behrend2011viability, hauser2016attentive}. 
Weighing the affordances and limitations, AMT participants are thus a reliable and accessible comparison group, one that reasonably satisfies our initial desire to create a high-contrast comparison between potential creators and end-users of AI systems. 

Our operationalization of the two groups' AI backgrounds aligns closely with the human-grounded evaluation proposed in~ \cite{doshi2017towards}, in which participants conduct controlled tasks to get a formative sense of the affordances of the explanations. We acknowledge that there are limitations to this experimental setup and that our insights should be scoped accordingly (more in Section~\ref{sec:limitations}). However, as we will see later, even with a carefully controlled task, we discover surprising, non-intuitive insights about how different groups interpret explanations.\footnote{\textit{Practices for better data}: We share some strategies that helped us gather high-quality data and maintain rapport with our participants throughout the study (e.g., fair payment structure, active engagement participants, etc.) in the Appendix (\ref{bestpractices}), uploaded with Supplementary Material.}

\subsubsection{Recruitment and Screening Methods} \label{sec:screening_and_recruit}
\hfill\\
All participants received US $\$10$ for their time. The \textbf{AI group} had 96 US-based participants (39\% self-identifying females; rest as males) taking $31.1$ minutes on average for task completion. We recruited undergraduate students enrolled in an AI course at a large public research university located in the US. 
Typically taken in the 3rd year, this is a keystone course in an AI degree specialization track, implying that a significant number of students have expressed longitudinal interest in AI. 
The course is taught using the most widely adopted textbook (by Russell \& Norvig~\cite{russell2010artificial} taught in over 1500 schools worldwide~\cite{aima_russell_norvig_bookadoption}) and using a widely adopted set of assignments.
While there is no guarantee that all students will be future AI creators, the faculty believes that we can reasonably assume that many students aspire to have careers in the development of AI technology.
In the course, students learn and implement many foundational AI concepts; for instance, Markov Decision Processes and Reinforcement Learning.
Our study was deployed after students had taken exams on these concepts. 
For the \textbf{Non-AI group,} we recruited participants from Amazon Mechanical Turk (MTurk) using TurkPrime~\cite{litman2017turkprime}. The non-AI group had 83 US-based participants (46\% self-identifying females; rest as males) taking $29.8$ minutes on average for task completion. 

\textbf{Establishing group differences while controlling for shared characteristics:} We systematically screened \textit{both} groups and ensured they are \textit{measurably} different. Every participant had to pass three parts of our screener: (1)~A knowledge test on CS \& AI (collaboratively developed with the course's teaching staff to ensure relevancy); (2)~ Self-reported knowledge levels in computer programming and AI using two 5-point Likert-scales; (3)~ confirmation of whether they have ever taken an AI class. The \textit{cut-off points} were decided through iterative piloting and consultation with the teaching staff of the AI course. The \textbf{non-AI group}, by design, should have ``No knowledge'' $[=1]$ in both programming and AI along with no prior AI classes. Out of 5, the \textbf{AI-group} should score $[>=4]$ (``moderate'' knowledge or more) for (1) [knowledge test]; for (2) [self-reported AI knowledge], it should be $[>=3]$ (``some knowledge'' or more). 

To establish that these \textit{two groups are measurably different}, we performed statistical tests. Mann Whitney U-test with Bonferroni correction showed that the groups are different--- for the AI knowledge test, $p <  2.2\times~10^{-16}$; for the self-reported programming knowledge, $p <  2.2\times~10^{-16}$. Thus, we were able to establish two ``high contrast'' groups. For further details, including screener questions, please refer to \ref{participant screener} in the Appendix, uploaded with Supplementary Materials.
Beyond the differences, we controlled for factors such as age and education levels. All participants were adults up to 25 years old, using smartphones and laptops every day. The non-AI group self-reported an average education level of $4.95$ (4= “Vocational Training”, 5= “Some college/Associate’s degree”) while the AI group's average was $5$. Mann Whitney U-tests showed that the groups did \textit{not} differ along \textit{age} ($p=0.505$) and \textit{education levels} ($p=0.146$).
\vspace{-10pt}

\subsection{Procedure: Task Details}\label{sec:task_details}
We now discuss study mechanics--- after providing informed consent, participants watched an orientation video outlining the scenario.
Participants were asked to imagine themselves as space explorers faced with a search-and-rescue mission involving robots.
They face a life-and-death situation where they are stuck on a different planet and must remain inside a protective dome. 
Their \textit{only} source of survival is a remote supply depot, which they cannot reach.
They must rely on autonomous robots, \textit{ones they cannot control}, to navigate through a field of boulders and a river of flowing lava to retrieve the essential food supplies~(See Fig~\ref{fig:3robots}). 
Participants could only see a non-interactive video stream of their activities through their "space visors". 
This non-interactiveness aimed to heighten their sense of lack of control (and thereby reliance on the robots).
Since the robots took identical actions during the task, participants were asked to pay special attention to the only differentiating factor-- the way each robot explained its actions. 

After orientation, participants watched 6 counterbalanced and randomized videos showing the three robots succeeding and failing to retrieve the essential supplies using identical sequences of actions.
To mitigate the effects of preconceived notions, we did not use any descriptive names for the robots; instead, we introduced the robots as “Robot A,” “Robot B,” and “Robot C” for the RG, AD, and NR robots respectively.
Participants, by design, had no idea about how each robot generated its expressions;  for instance, participants were not informed that NR's numbers are Q-values. To them, it was just another different way of explaining the actions. As mentioned in \ref{sec:explanation_gen_description}, the goal was to see how and if how people can make sense of them even if they are unlabelled.
After watching all the videos, participants ranked the robots (1st, 2nd, and 3rd-- no ties allowed) along the five (5) dimensions of user perceptions highlighted above (in \ref{sec:user_perceptions}). 
After ranking on a dimension, participants justified and contextualized their ranking using a mandatory free-text response.

\section{Quantitative Results}\label{sec:quant_analysis}

We conducted within-group and between-group analyses.
The \textit{within}-group analyses give a sense of which robot's explanation was preferred along each dimension of user perception.
The \textit{between}-group analyses tell us how the two groups are the same or different when considering the robots across multiple dimensions and set the stage for the qualitative analysis in Section 5. Taken together, the quantitative results address RQ1 around quantifying the relative perceptions and preferences of the two groups along the five dimensions.

\subsection{Within-group Comparisons}
\label{sec:quant-within}

\begin{figure*}[tbh]
    \vspace{-4mm}
    \caption{Distributions of rankings for each robot, in each dimension, separated by participant group}
    \centering
    \includegraphics[width=10cm]{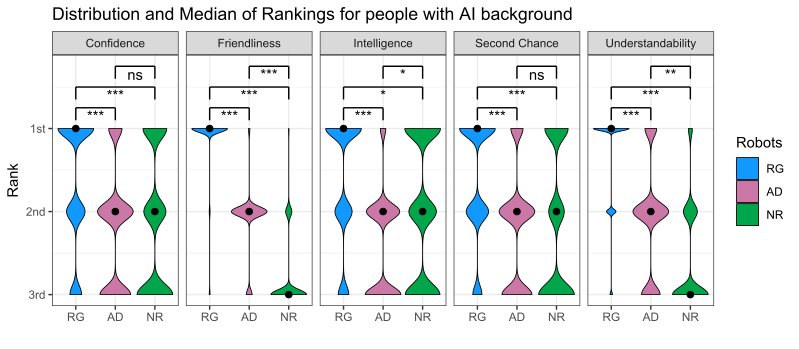}
    \includegraphics[width=10cm]{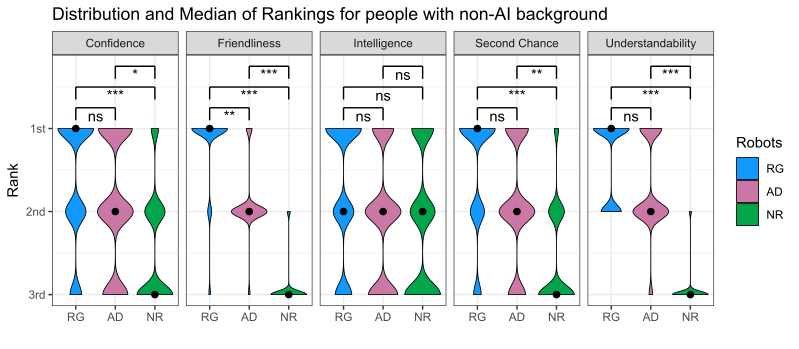}
    \begin{fignote}
    Note: The black horizontal bars indicate whether a pair of distributions is significantly different. ns = not significant, *$p < .05$, **$p < .01$, ***p < .001. The width of each violin plot at each ranking level indicates the proportion of people who assigned that rank to that robot. The black bullet ($\scriptstyle \bullet$) refers to the median rank.
    \end{fignote}
    \label{fig:violins}
    \Description[Violin charts for AI-group and Non-AI group, along each dimension.]{Two rows of five violin charts. The top row shows AI-group results for the confidence, friendliness, intelligence, second chance, and understandability dimensions. The second row shows Non-AI-group results for the same dimensions. Each of the 10 violin charts shows three vertical bars, one for each robot (RG, AD, and NR), with bars wider in the top, middle, or bottom based on whether the respective group ranked each robot more often as their first, second, or third pick for the dimension.}
\end{figure*}

Figure~\ref{fig:violins} shows how each robot was ranked in each dimension.
The ``violin chart'' visualization makes similarities and differences easier to distinguish. 
For example, large differences are exemplified in the \textit{Friendliness} rankings given by the non-AI group participants (bottom, second left). 
A wide area at the top of the blue plot shows that the Rational Generating (RG) robot was ranked first by most.  
In contrast, the plot for the \textit{Intelligence} dimension by the non-AI group (bottom, middle) shows that all robots received the first, second, or third rank comparably similar number of times. 

For each group, we conducted five Friedman tests~\cite{zeileis2008implementing} of differences among repeated measures (one for each of the five dimensions). 
We used the maximum-type (max T) implementation of the Friedman test, which controls for the family-wise error rate \cite{zeileis2008implementing} to determine {\em whether} any preferences between robots were detectable.
To determine {\em which} robot was preferred, we used the Wilcoxon-Nemenyi-McDonald-Thompson test~\cite{hollander2013nonparametric} to make pairwise comparisons between the robots, for each participant group, and within each dimension.
The results are summarized in Table ~\ref{table:pairwise}.

\begin{table*}[tbh]
\caption{Summary of $p$-values for pairwise comparisons, showing which robots were preferred.} 
\label{table:pairwise}
\centering
\begin{tabular}{llrllr} 
\toprule
Dimension         & \multicolumn{2}{c}{{\cellcolor[rgb]{1,0.8,0.431}}AI background} &  & \multicolumn{2}{c}{{\cellcolor[rgb]{1,0.8,0.431}}Non- AI background}  \\
                  & \multicolumn{1}{c}{Robot} & ~ ~ ~$p$-value             &  & \multicolumn{1}{c}{Robot} & ~ ~ ~$p$-value                   \\ 
\cline{1-3}\cline{5-6}
                  & \textbf{RG} vs AD & \textbf{< 0.001}                              &  & RG vs AD          & 0.362                                             \\
\textit{Confidence}        & \textbf{RG} vs NR & \textbf{< 0.001}                              &  & \textbf{RG} vs NR & \textbf{< 0.001}                                    \\
                  & AD vs NR          & 0.869                                       &  & \textbf{AD} vs NR & \textbf{0.014}                                    \\ 
\cline{1-3}\cline{5-6}
                  & \textbf{RG} vs AD & \textbf{< 0.001}                              &  & \textbf{RG}~vs AD & \textbf{< 0.001}                                    \\
\textit{Friendliness}      & \textbf{RG }vs NR & \textbf{< 0.001}                              &  & \textbf{RG~}vs NR & \textbf{< 0.001}                                    \\
                  & \textbf{AD} vs NR & \textbf{< 0.001}                              &  & \textbf{AD}~vs NR & \textbf{< 0.001}                                    \\ 
\cline{1-3}\cline{5-6}
                  & \textbf{RG}~vs AD & \textbf{< 0.001}                              &  & RG vs AD          & -                                                 \\
\textit{Intelligence}      & \textbf{RG} vs NR & \textbf{0.021}                              &  & RG vs NR          & -                                                 \\
                  & AD vs \textbf{NR} & \textbf{0.011}                              &  & AD vs NR          & -                                                 \\ 
\cline{1-3}\cline{5-6}
                  & \textbf{RG}~vs AD & \textbf{< 0.001}                              &  & RG vs AD          & 0.083                                             \\
\textit{Second Chance}     & \textbf{RG} vs NR & \textbf{< 0.001}                              &  & \textbf{RG~}vs NR & \textbf{< 0.001}                                    \\
                  & AD vs NR          & 0.902                                       &  & \textbf{AD}~vs NR & \textbf{0.003}                                    \\ 
\cline{1-3}\cline{5-6}
                  & \textbf{RG}~vs AD & \textbf{< 0.001}                              &  & RG vs AD          & 0.187                                             \\
\textit{Understandability} & \textbf{RG} vs NR & \textbf{< 0.001}                              &  & \textbf{RG~}vs NR & \textbf{< 0.001}                                    \\
                  & \textbf{AD}~vs NR & \textbf{0.002}                              &  & \textbf{AD}~vs NR & \textbf{< 0.001}                                    \\
\bottomrule
\end{tabular}
\begin{fignote}
\centering
    Note: Favored robot and significant $p$-values are in bold for each pairwise comparison.
\end{fignote}
\Description[A table showing $p$-values for each dimension and robot and pairwise robot comparison.]{The table has five rows, one for each dimension: confidence, friendliness, intelligence, second chance, and understandability. The first column is for the AI background group and shows each pairwise robot vs. robot comparison (e.g., RG vs. AD) and the $p$-value indicating whether the rankings are significantly different between the two robots and for the participant group and the dimension.}
\end{table*}

From the within-group analysis (Table ~\ref{table:pairwise} and Figure~\ref{fig:violins}), we have some notable insights: across each dimension, the AI-background group unambiguously preferred RG to the other robots, particularly over AD.
However, the RG robot is not the unanimous winner across each dimension for the non-AI group-- here, participants show no preference between RG and AD in 4 out of the 5 dimensions (RG wins over AD in {\em Friendliness}). 
For the non-AI group, AD wins over NR across all dimensions except for {\em Intelligence}, which is a noteworthy dimension--the non-AI group showed no preferences between the robots, but the AI group felt NR was more intelligent than AD. This is the only time where NR wins over AD, highlighting an important point that we explore in our qualitative findings (Section \ref{sec:faith_num} and \ref{sec:explanatory_value}) around the AI group's preference for numerical representations.  

\subsection{Between-group Comparisons} \label{sec:quant_between_groups}
To detect group differences in the preference for the robots, we used Ordinal Logistic Regression (OLR) \cite{mccullagh1980regression, agresti2003categorical,R_MASS,OLR_UCLA}, an extension of Logistic Regression when the response variable is ordinal.
To model a 3-level categorical variable~(Robot Type), OLR requires us to analyze two variables holding one as a reference (constant): in our case, we analyzed AD and NR, holding RG constant.
We investigate the interaction effects as well as changes in reference levels in the OLR analysis that reveals the relative impact in ranking the robots between the groups. 
To investigate the effect of the dimensions, we explore the interaction effects between Robot Type (RG, AD, and NR) and the Participant Group (AI vs. non-AI) into the OLR model. 
Changing the reference levels of Robot Type and Participant Group allows us to isolate the interaction effects, which we interpret similarly to~\cite{james2001interaction}.
A full list of OLR tables is provided in the Appendix (Table~\ref{tab:ANon-Turk}-~\ref{tab:CTurk}).
If we group everyone regardless of their AI backgrounds, 
we find that $Rank_\text{RG}>Rank_\text{AD}>Rank_\text{NR}$. The odds of receiving a higher ranking for the RG robot is 5.5 times that of the AD robot, whose odds are 2.66 times that of the NR robot (see Tables~\ref{tab:Overall_Robot_Ranking_RG baseline} and~\ref{tab:Overall_Robot_Ranking_AD baseline} in the Appendix (\ref{OLR tables})).

\textit{Between the groups}, there is no significant difference in ranking RG Robot---it is always the top choice across all dimensions~(Table \ref{tab:each_p_summary}, top row). 
However, the groups exhibit differences in their preferences when it comes to AD and NR. 
The Non-AI group shows more preference for the AD Robot (odds ratio= $1.986$) while the AI group shows more preference for the NR Robot (odds ratio= $1.0/0.465=2.15$). 

Table \ref{tab:each_p_summary} also provides the odds ratio and the $p$-value for each Robot Type per dimension. 
For the RG Robot, all of its $p$-values are greater than 0.05 for each of the five dimensions resulting in no significant pattern of preference between the groups.
Conversely, the $p$-values of AD Robot are all smaller than 0.05 
meaning that, when it comes to ranking the AD Robot, there is a significant difference between the Non-AI group and the AI group.
Moreover, all odds ratios related to AD Robot in Table \ref{tab:each_p_summary} are greater than $1.0$, indicating the Non-AI group shows a stronger preference for the AD Robot than the AI group for each of the five dimensions. 

Last, for the NR robot, the $p$-values related to \textit{Confidence}, \textit{Second Chance}, and \textit{Understandability} are also significant.
The odds ratios for these dimensions are less than $1.0$ indicating that the AI-group is more likely to rank the NR robot higher than the non-AI-group on these dimensions.
There is no significant difference between the Non-AI group and the AI group when it comes to ranking the dimensions of \textit{Friendliness} and \textit{Intelligence}.

\begin{table*}[tbh]
  \caption{Robot Preference Summary across Dimensions (Non-AI [baseline] vs. AI) 
  \label{tab:each_p_summary}}
  \centering
  \begin{tabular}[t]{lllllll}
  \toprule
     & \multicolumn{2}{c}{RG Robot} & \multicolumn{2}{c}{AD Robot} & \multicolumn{2}{c}{NR Robot} \\
     \cmidrule(l){2-3}\cmidrule(l){4-5}\cmidrule(l){6-7}
     Dimension& Odds ratio & $p$-value & Odds ratio & $p$-value & Odds ratio & $p$-value \\
\midrule
     All dimensions & 0.855 & 0.323 & 1.986  & $\boldsymbol{<0.001}$ & 0.465 & $\boldsymbol{<0.001}$\\
  \midrule
     Confidence & 0.971 & 0.930 & 2.003 & \textbf{0.026} & 0.487 & \textbf{0.029}  \\
     Friendliness & 0.280 & 0.051 & 2.827 & \textbf{0.017} & 0.436 & 0.160 \\
     Intelligence & 0.717& 0.315 & 1.959 & \textbf{0.031} & 0.646 & 0.178 \\
     Second Chance & 1.369 & 0.356 & 1.965 & \textbf{0.028} & 0.375 & \textbf{0.005 } \\
     Understandability & 0.659& 0.255& 3.088& \textbf{0.005} & 0.114 & \textbf{0.006 } \\
  \bottomrule
\end{tabular}
 \begin{fignote}
 \centering
   Note: Odds ratios $>1.0$ indicate the non-AI group prefers the robot more than the AI group.
  \end{fignote}
  \Description[A table summarizing the preferences for each robot across all dimensions, comparing AI to Non-AI groups]{The table has six rows: all dimensions, confidence, friendliness, intelligence, second chance, and understandability. For each row, the first two columns show the odds ratio and the $p$-value for the RG robot. The third and fourth columns show the odds-ratios and $p$-values for the AD robot. The last two columns show the odds-ratios and $p$-values for the NR robot.}
\end{table*}

\subsection{Conclusions of Quantitative Analyses}

Overall, while we find that $Rank_\text{RG}>Rank_\text{AD}>Rank_\text{NR}$, there are significant differences in the AI and the Non-AI ranking behaviors.
In other words: AI background does change perceptions of explanations along the dimensions we identified.
In particular, having a background in AI correlates with having a greater preference for the NR Robot over the AD Robot.
While AD wins over NR in most head-to-head comparisons, NR wins it on {\em Intelligence} for the AI group.
The two groups are indistinguishable concerning {\em Friendliness}.
The next section helps us understand why these trends are present in the quantitative results and their implications.

\section{Qualitative Analysis \& Findings}\label{sec:qual_findings}

While quantitative analysis tells us \textit{what} is different between groups, qualitative analysis sheds light on \textit{how} and \textit{why} those differences manifest, addressing RQ2.
Quantitative analysis puts RG as the clear winner; however, as we show in this section, the story is more nuanced and interesting, especially the underlying reasons behind the group differences between AD and NR.
Below we describe the process of our qualitative analysis before moving on to the findings. 

The qualitative data was coded according to principles of \textit{grounded theory }analysis \cite{Charmaz2014, Strauss1990}.
Coding and analysis of the qualitative data were done by the first and second authors in two stages, each consisting of multiple rounds of iteration.
In the first round, the authors separately performed coding using \textit{in-vivo} codes, which involves generating codes from the data (e.g., using participant phrases such as `the robot knew what it was doing' and `easier to understand' as codes).
Through discussion, the authors generated a merged open coding scheme.
In the second round, the authors analyzed the data using \textit{axial codes}.
Axial coding involves finding connections between open codes and classifying them into different categories (e.g., potential actionability of numbers).
In the last step, the authors analyzed the different sets of axial codes, unified them into high-level themes, and consolidated them into selective codes.

After this, we grouped the data along the two groups (AI vs. non-AI) and our five~(5)~analytic perception dimensions that we motivated in Section~\ref{sec:user_perceptions} and also used for quantitative analysis~(\textit{confidence, friendliness, intelligence, second chance,} and \textit{understandability}), allowing us to compare our findings across dimensions and groups.
Finally, based on the grounded theory heuristic of ``constant comparison'' \cite{Glaser1967, Strauss1990}, the authors frequently compared and contrasted axial codes across perception dimensions to tease out the similarities and differences between the different reasonings used by the AI and non-AI groups to make sense of the explanations provided by the three robots.

Below, we report on the most salient themes (selective codes) from our analysis.
In particular, we showcase \textbf{(1)}~how irrespective of their AI background both groups exhibited unwarranted faith in numbers for different reasons and
\textbf{(2)}~how each group saw explanatory value in explanations that were not designed with justificatory qualities.
For each theme, we highlight distinct categories of reasoning (axial codes) used by the groups to justify their choice of robots across the different perception dimensions.

\subsection{Unwarranted Faith in Numbers}\label{sec:faith_num}

\textbf{Participants in both groups had unwarranted faith in numbers. However, their \textit{extent} and \textit{reasons} for doing so were \textit{different}}. 
On the one hand, AI group participants often ascribed more value to numbers than was justified. On the other hand, some non-AI group participants believed that numbers signaled intelligence even if participants could not capture the numbers' meaning. Below we highlight two major ways in which participants misplaced faith in numbers.

\textbf{The mere \textit{presence of numbers} was associated with an \textit{algorithmic thinking process} in the robot even when the meaning of numbers was unclear}. 
Both groups exhibit this perception of algorithmic thinking. We will begin with the AI group---this group ascribed higher-order cognitive abilities to the robot with numerical representations. Between AD and NR, the AI group found AD more understandable (Table~\ref{table:pairwise}) but deemed the NR robot as the more intelligent of the two (the only time NR wins over AD across all dimensions-- (Table~\ref{table:pairwise})).
It seems contradictory that the AI group found the \textit{less understandable} robot to be \textit{more intelligent!}---the main reason for this is that the presence of numbers fostered the “assumption that [the NR robot] uses some sort of [an] algorithm” (A50) in the AI group. The perception of “under-the-hood math, boosted [NR’s] trustworthiness” (A49).  
Some explicitly compared AD and NR robots and concluded that the mathematical representation demonstrated a method for the NR robot’s behavior:

\begin{quote}
    “With [the NR robot], while I did not understand its methodology, I could see that it was using some mathematical calculations to determine which way to move. […] With [the AD robot], I could not see any methodology or signs of decision-making” (A23).
\end{quote}

A small minority in the AI group indeed figured out that “the numbers 0-4 represented different actions, and [NR] would choose the action with the highest numerical value” (A23). An even smaller minority even guessed the numbers’ correct meaning—a “utility function and reward systems” (A64).
\textit{despite} the fact that the numbers were \textit{unlabelled}. These participants projected meaning on the numbers, lending further credence to the role of data vision~\cite{Goodwin1994, Passi2017}.

What is surprising is their faith in numbers arose even when AI group participants did not “fully understand the logic behind [...NR robot’s] decision making.'' (A43). \textbf{The AI group seems to have followed heuristic reasoning that associates mathematical representations with logic and intelligence}, e.g.: "logic must have been derived from a formula, [...which is] intelligent'' (A54), or ``Math [...had] an aura of intelligence" and ``exact values” made the NR robot “feel smarter” (A16, A77, A75). This perception of logical thinking engendered unwarranted trust and made the NR robot seem like it "should theoretically succeed more than the others, making him more intelligent” (A37). 
The linkage between perceived logical thinking and higher cognitive abilities can elucidate why the AI group prefers NR over AD when it comes to \textit{Intelligence} (Table~\ref{table:pairwise}).
A few participants even claimed that they could “actually see the math that [the NR robot] was making decisions off of, [...making it feel] more real” (A9). In fact, to them, it appeared as if the NR robot “\textit{clearly had an algorithm} that worked […and]\textit{ seemed to know }what it was doing” even if they “did not know what it was going to do in the future” (A91, emphasis added).

\textbf{
Participants with AI background also viewed numbers as \textit{potentially actionable} even when their meaning was unclear}. \textit{Actionability} refers to what one might do with the information to make sense of the robot’s behavior---``debug its faulty behavior, or predict its future behavior'' (A76). While many highlighted that they could not ``make sense of numbers right now, [they believed that] in principle, [they] should be able to act on them in the future'' (A39).  The potential explanatory value in numbers was better than the vacuous statements of AD: “[The NR] robot gave mathematical results as explanation while [the AD robot] gave no explanation” (A19).

Many participants connected their AI background to their ability to work with numbers. 
They mentioned how their current “AI course [helped them] to understand what [NR robot] is doing and what the numbers might mean” (A52). Some even highlighted that if they “had a pen and some paper, writing down information, [they] could gleam [sic] some information based off the [numerical] patterns” (A28). 
But how \textit{actionable} were NR’s numbers in actuality? 
The numbers were Q-values, which only indicate the relative strength of one action versus the others.
Specifically, Q-values indicate the agent's belief that certain actions lead to greater or lesser future reward, affording some amount of explanatory power.
However, they cannot indicate {\em why} the agent has come to believe that one action is better than another---this information is not retained by the agent nor conveyed through the numbers. 
This did not deter the AI group from deferring to the authority of numbers. After all, while all the robots used the same AI algorithm to make decisions, the NR robot’s expressions seemed most `AI-like'. 

These insights can help us understand why, even though both groups have misplaced faith in numbers, the AI group shows a higher inclination towards numbers (as evidenced in the quantitative results (Section~\ref{sec:quant-within}, Table~\ref{table:pairwise}), the only time NR wins over AD happens when the AI group judges \textit{Intelligence}).

\textbf{Unable to access their meaning, the non-AI group associated numbers with the presence of a higher, more intelligent expression} that, they argued, could only come from an intelligent agent. Since the NR robot was “communicating in a numerical language that's too hard to understand”, the numbers had a "mystery and aura of higher intelligence"~(NA22, NA33). The “language of numbers”, \textit{because} of its “cryptic incomprehensibility,” signalled higher-order thinking (NA6, NA1): “\textit{Because} I could not understand [NR's] output, I deemed it to be intelligent” (NA30, emphasis added).

This can explain why the non-AI group showed no preferences between AD and NR when judging \textit{Intelligence}, despite favoring AD over NR across all other dimensions~(Table~\ref{table:pairwise}).
To them, numbers signaled ``precision''---an important quality of intelligence (NA8, 16, 21, 30, 34). Numbers gave the impression that the NR robot “was more technical” than others—its “precise numerical explanations” resulted from it “calculating everything” (NA53, NA21, NA12). To these participants, “anything that uses pure numbers is going to be more intelligent” because “numerical outputs are likely to be more precise […] whereas textual representations involve a degree of uncertainty and subjectivity” (NA41, NA30). Such perceptions point to how the modality of expression—numeric vs. textual—impacts perceptions of explanations from AI agents, where we see projections of normative notions (e.g., objective vs. subjective) in judging intelligence.

\subsection{Unanticipated Explanatory Value}\label{sec:explanatory_value}

As designers, we had specific goals behind each robot's mode of expression. As discussed in Section \ref{sec:explanation_gen_description}, RG was the only robot designed to have the justificatory quality to explain the \textit{why} behind the robot's actions, while AD and NR (as baselines) should be considered as lacking justificatory qualities by-design. 
RG won the overall competition (as discussed in Section \ref{sec:quant_between_groups}), but what was surprising was that \textbf{both AI and non-AI groups found unanticipated explanatory value in AD's declarative statements and NR's numerical representations.} 
As we will show below, qualitative analysis revealed that the two groups had different \textit{explanatory intent} (what they wanted to do with the explanation), which was closely associated with their AI background. On the one hand, the non-AI group found \textit{affirmatory} value (confirmation of stable performance) in AD's statements. On the other hand, the AI group overly ascribed \textit{diagnostic} value (debugging in case of failure) to NR's numbers even when they could not make sense of them.

For the non-AI group, their \textbf{desire for \textit{affirmation}} --- confirming the action without necessarily explaining it--- played a key part in finding value in AD's explanations. The affirmatory value manifests most clearly in their comparison between AD and NR in the dimensions of \textit{Confidence} and \textit{Understandability}. For both dimensions, the non-AI group preferred AD over NR~(Table~\ref{table:pairwise}). Recall that AD merely declared its action, stating the \textit{what}, not the why. Despite this, the non-AI group found value in the confirmatory information because there was alignment between what AD was doing and saying. It showed that ``[AD] is consistent and nothing crazy is going on where it says it went right but in actuality, it went down'' (NA34). For both dimensions, the non-AI group attributed greater value to AD’s declarative statements (compared to NR's numbers) because it ``at least said what its movements were going to be" (NA7). Its “brief,” “un-embellished," and “easier to understand" language that got "straight to the point" boosted its \textit{understandability} (NA14, NA23, NA28, NA7). In fact, AD's ``just the facts'' (NA38) declarative and succinct nature were signs of \textit{confidence} itself. It did not need to say much because ``it knew what it was doing,''---evident in its lack of ``any hesitation" and "business-like [style] focused on performing the task at hand" (NA17, NA41, NA11). In contrast, NR's numbers were inaccessible to the non-AI group, thereby in-actionable and valueless.
In short, when we designed AD, there was no intention of offering value through confirmation; yet, \textbf{through their explanatory intent of affirmation, the non-AI group found value in AD's explanations by interpreting them as signals for stable system performance}.

\textbf{The AI group overly ascribed \textit{diagnostic} value in NR's numbers even when their meaning was unclear}---they felt NR's numbers had “diagnostic information that can be used to debug [the robot] in case of failure” (A39). The explanatory intent of \textit{diagnosis} is a consequence of the AI group's faith in numbers and related to their perceived actionability---what one might do with the information (as discussed above in~\ref{sec:faith_num}). 
Analysis of the open-ended text responses for the perception dimensions of \textit{Intelligence}, \textit{Second Chance}, and \textit{Confidence} exhibits this explanatory intent most clearly.
Recall that the AI group preferred NR over AD for Intelligence and felt there were no differences between them for Second Chance and Confidence~(Table~\ref{table:pairwise}). This is in contrast to the non-AI group that preferred AD over NR for both of these dimensions. 
AI group members perceived NR’s numbers to have more explanatory value simply because they felt they could do ``more with numbers'' (A30) in ``cases of failure and troubleshooting''(A78).

For Intelligence, NR's numerical representation and   "exact values" made it appear more ``valuable'' than AD's ``inert'' statements (A23, A51, A42) . 
Even when it came to giving NR a second chance, numbers helped because the NR robot appeared to be “trying very hard since it provided [...] mathematical evidence of its moves” (A79).“Math-based decisions” (A2) made the NR robot appear “more reliable” (A8, A42) and worthy of another chance.
Numbers inspired confidence because they signaled the existence of a "concrete methodology"—--"the higher the number, the more optimal the move" (A66, A67), which added to the perceived explanatory value in them. 
Moreover, if there were a method, an algorithm, or a formula behind the NR’s actions, the AI group participants believed that they could deduce it using the numbers:

\begin{quote}
    "[NR] returned the most amount of data, and \textit{if I were to understand what that numbers mean}, it'd be the most useful one to debug on repeating runs to analyze what it is doing and why" (A91, emphasis added).
\end{quote}

The AI group participants connected their background to a desire to troubleshoot and repair things. Numbers, they felt, were more manipulable than language, catalyzing over-ascription of diagnostic value to them:

\begin{quote}
    ``As an engineer, I want to fix things. Numbers are concrete and objective, language is not. But you can manipulate numbers [...] With the AI stuff I'm learning, I can use it to diagnose [NR].'' (A49) 
\end{quote}

Thus, we see, what Eriksson called the "seductive allure" \cite{Eriksson2012, LOMBROZOpreferences} exhibited through the perception of numbers in the AI group. This diagnostic intent showcases how the AI group over-ascribed explanatory value in numbers, highlighting how its unwarranted faith in numbers can manifest in potentially negative ways---a point we discuss in Sections~\ref{sec:discussions_and_implications}.

In \textbf{summary}, \textit{first}, both groups had undue faith in numbers for \textit{different reasons} and to \textit{different extents}. Both groups associated the presence of numbers to signal the presence of algorithmic thinking even when the meaning of numbers was unclear. The AI group seems to have employed heuristic reasoning, connecting mathematical representations with the notions of logic and intelligence. The non-AI group, lacking access to their meaning, correlated numbers with the existence of a higher, more intelligent expression. 
\textit{Second,} both AI and non-AI groups found unanticipated explanatory value by appropriating AD's declarative statements and NR's numerical representations. The non-AI group had an explanatory intent of \textit{affirmation} (confirming the action without explaining it). This enabled them to find value in AD's explanations by interpreting them as signals for stable system performance. Notably, the AD robot was not designed to offer value through confirmation. The AI group overly ascribed \textit{diagnostic} value in NR's numbers for potential debugging even when their meaning was unclear. The explanatory intent of \textit{diagnosis} is a consequence of the AI group's unwarranted faith in numbers and related to their perceived actionability---what one might do with the information (as discussed above in~\ref{sec:faith_num}).

\section{Discussion \& Implications}\label{sec:discussions_and_implications}

In this section, we discuss possible causes of the observed differences between the groups, then discuss implications for designing XAI systems by accounting for users' background differences to bridge the AI creator-consumer gaps, and then the broader implications for explainable and responsible AI.

\subsection{Discussion: User background impacting perception of XAI}\label{sec:heuristcs_appropriation_6.1}

We \textbf{highlight two ways to interpret the potential causes behind group differences-- cognitive heuristics and appropriation}. First, we use the notion of \textit{heuristics} based on the dual-process theory (reviewed in Section~\ref{background_hcxai}) to understand how unwarranted faith in numbers (Section~\ref{sec:faith_num}) can emerge from different lines of thinking affected by one’s AI background. Second, we incorporate the lens of \textit{appropriation} to the unanticipated ways in which people find explanatory value (Section~\ref{sec:explanatory_value}).

\vspace*{-3mm}
\paragraph{Heuristics and Faith in Numbers:}

Recent work (reviewed in ~\ref{background_hcxai} like ~\cite{buccinca2020proxy,ghai2021explainable}) has highlighted that while XAI is often developed with an implicit assumption that the recipient will process each explanation through analytical System 2 thinking (from dual-process theory~\cite{kahneman2011thinking}), in reality people are more likely to rely on System 1 thinking by invoking \textit{heuristics}--rules-of-thumb or mental shortcuts, which leads to biases and errors if applied inappropriately~\cite{kahneman2011thinking,wason1974dual,petty1986elaboration}.

The notion of \textit{heuristics} can help us understand the potential reasons behind the two groups' different faith in numbers.
On the one hand, the AI group seemed to have an instinctual response to numerical values; they assumed that the numbers possessed all the information needed to manipulate, diagnose, and reverse engineer. There appear to be heuristics
that associate numbers with logical intelligence, which could potentially be acted upon (e.g., through diagnosis). 
Such heuristics are likely formed and validated from their past experience working with numbers and algorithms. This heuristic is risky because, as we noted in Section \ref{sec:explanation_gen_description}, the numbers are Q-values and do not allow much actionability beyond an assessment of the quality of the actions available. We return to this risk and highlight design implications to address this in the next subsection.

On the other hand, lacking the AI background, some in the non-AI group seemed to have different heuristics where their very inability to understand complex numbers was associated with the presence of a higher-order intelligence.
The non-AI group also lacks the requisite AI background to potentially access and deliberatively think through NR's numbers (System 2 thinking).
Thus, we see how different heuristics, tied to one’s AI background, can lead people to the same outcome—faith in numbers. Note that these heuristics are not an exhaustive list, but a starting point to understand how group differences connect to their AI backgrounds. 

\vspace{-3mm}
\paragraph{Appropriation and Unexpected Explanatory Value:}A second lens for understanding why participants found explanatory value in unexpected places is that of {\em appropriation}.
As we shared in Sections \ref{sec:individual-differences} and \ref{sec:explanatory_value}, appropriation happens when end-users interpret and use technologies in ways not envisioned by designers \cite{dix2007designing, muller2016collaborative, salovaara2008inventing, tscheligi2014potentials}. People do not passively absorb information—they interpret it, often processing it in unanticipated ways. This is what happened to AD’s declarative statements and NR’s numbers. Driven by different explanatory intents, each group appropriated the explanations in unanticipated ways. These different appropriations were, in part, a function of each group's AI background that led participants to develop their own sense of how they can and cannot use explanations. What is striking is that the appropriation took place even in a controlled experiment like ours where the participants were not explicitly asked to take actions based on explanations. Even in passive interactions (robots engaged in one-way communication), participants envisioned themselves \textit{using} the explanations in hypothetical scenarios. On the one hand, the AI group’s intent of \textit{diagnosis} led many to envision scenarios where they would troubleshoot the robot. As a result, they might have misplaced or over-placed diagnostic value in NR’s numbers—even when they could not fully understand them. On the other hand, for the non-AI group, NR’s numbers are inaccessible, thus in-actionable. Their \textit{affirmatory} intent drives them to find value in the confirmatory statements of AD, appropriating it as a signal for stable system behavior.

\textbf{Our discussion extends the literature around individual differences and heuristics-based processing of XAI~\cite{ghai2021explainable,nourani2021anchoring,buccinca2020proxy} in two ways}: first, we explicitly highlight \textit{what} heuristics people might use, \textit{how} their AI background influences their thinking, and posit the \textit{why} (underlying reasons) behind them. Second, we add the lenses of appropriation to the conversation, which has implications on user agency in design (points we touch on below). For both, we connect it to an important user characteristic—one’s AI background. 

Both points around heuristics and appropriation highlight an important yet overlooked point around the \textit{duality} of explanations. Explanations are \textit{both products} and \textit{processes} ~\cite{lombrozo2011instrumental}. The product-centered view, common in psychology and XAI, often ignores the essential processes through which people make sense of explanations. However, the sensemaking process is as important as the explanation itself~\cite{lombrozo2012explanation, wilkenfeld2015inference}. Recent work calls for guiding the process of understanding explanations, especially for non-AI experts~\cite{chromik2021think}.  By investigating both AI and non-AI groups, our work extends the current XAI discourse-- it directly speaks to the duality of explanations by focusing on both products (types of explanations from 3 robots) and processes (how one’s AI background influences the interpretation of explanations).

\textit{Taken together, these findings reveal that the design and use of AI explanations is as much in the eye of the beholder as it is in the minds of the designer}--- the user's explanatory intent and common heuristics matter just as much as the designer's intended goal. Users might find explanatory value where designers never intended to be and use them based on their explanatory intent. Contextually understanding the misalignment between designer goals and user intent is key to fostering effective human-AI collaboration, especially in XAI systems \cite{Gabriel_2020, mohseni2020goals}. 
We demonstrated how different groups find meaning in different places---even when the meaning is misplaced. The ‘ability’ in explain-\textit{ability} depends on \textit{who} is looking at it and emerges from the meaning-making process between humans and explanations.

\textbf{While the groups differed in other aspects, both preferred the RG robot because of its \textit{command of language} exhibited through explanatory power (depth of reasoning) and variety (style and length).} RG appeared to have a ``good command of English deep enough to get at the why'' (A29). Moreover, they found RG's communication to be \textit{relatable} and exhibited a \textit{personality} even if the interaction was passive. It appeared to “include you in its thought process,” making you feel “as if you could talk to it” (NA11, A46). With RG robot, many felt “like you are having a conversation with a friend, analyzing some problem, and making the best decision for this problem.” (NA36) As RG navigated the terrain, its expressions included phrases like `I'm not just winning at life, I'm biwinning!', which participants found humorous and engaging. 
Since participants felt RG was ``the most humanlike in its explanations'' (A35), they attributed \textit{emotional intelligence} to it, which garnered empathy and support. Most people ``rooted for [the RG robot because they] liked its personality, felt connected to it, and wanted it to succeed.'' (NA27). 

Given our findings, it might be tempting to conclude that RG-style explanations should be used at all times. This is where the \textit{who} in XAI comes in--- the \textit{who} scopes \textit{how} to show the \textit{why}. If the explainability needs of the user are to achieve \textit{functional understanding}~\cite{lombrozo2014explanation}, then RG-style explanations are strong contenders. RG-style explanations are also suitable if the goal is to make AI explanations accessible to non-AI experts~\cite{ehsan2019automated}. However, if the goal is to provide a \textit{mechanistic understanding}~\cite{lombrozo2019mechanistic,paez2019pragmatic}, then RG-style explanations will not be actionable. Mechanistic understanding can be useful for debugging purposes. Here, a hybrid style could work where the functional understanding of RG is contextualized with the mechanistic affordances from Q-values in NR. Imagine an explanation format where the user can ``unfold'' the associated Q-values and ground the natural language explanation. There are engineering and resource costs for RG-style explanations. They need to be trained with human explanations, which can be a challenging data collection exercise for more complex environments. 
The full treatise of explanation design is beyond the scope of this paper; however, focusing on the needs and characteristics of the ``who''  is a good starting point.

\vspace{-10pt}

\subsection{Implications: Designing XAI for background differences}\label{sec:mitigation strategies_6.2}
Here, we discuss the implications for designing XAI systems that accommodate users' AI background differences. We focus on mitigating the risks of over-reliance on numbers which can potentially lead to negative consequences such as over-trust on AI systems ~\cite{passi2022overreliance}. 
Moreover, as those in the AI group are likely to be on the creation end while those in the non-AI group are likely to be towards the consumption and interaction end, our results have implications for bridging the AI creator-consumer gaps by bringing conscious awareness of the group differences to the design of XAI systems.  

Our discussion around \textbf{heuristics} carries design implications for both groups. We noticed how the AI and non-AI groups utilize different heuristics (mental short-cuts) in System 1 (fast, automatic) thinking to ascribe misplaced faith in them. Shifting people’s thinking from System 1 to System 2 (slow, deliberative) is not only an active area of research but is also a challenging one~\cite{chen2003exploratory,lambe2016dual}. 
There can be two actionable ways to tackle biases resulting from cognitive heuristics— first, locally \textit{at the time} of decision-making, we can use Cognitive Forcing Functions (CFFs)~\cite{croskerry2003cognitive} (prompts, delays, etc.) that can interject the heuristic reasoning, potentially allowing the person to engage in deliberative analytical thinking. Second, globally for future decision-making, we can utilize metacognitive strategies, often called cognitive forcing strategies, that include simulation training, increasing awareness of potential pitfalls of heuristics, etc.

To potentially prevent over-trust in numbers, we can do design interventions at the local and global levels. At a local design level, we can introduce CFFs that can break instinctive thinking patterns and promote mindful ones. What if the AI group members were prompted to reflect on their instinctive thinking through a combination of prompts and multi-modal explanations (blend between RG and NR)? 
Situating numbers in the context of language and vice versa can act as a CFF that could prompt the AI group members to reflect deliberatively (using System 2) and realize the limited nature of the numbers. At a global level, we can introduce simulation training that provides counterfactual (what-if) scenarios highlighting cases where numbers from the robots are erroneous or faulty (vs. correct Q-values).
For the non-AI group members who associate the opacity of numbers with higher intelligence, we can introduce scenario-based examples that explicitly highlight how indecipherable numbers can also be gibberish and useless. The goal here is \textit{not} to eliminate heuristic reasoning but to mitigate blind faith in a certain modality of explanation. Exposure to these scenarios can facilitate long-term calibration of trust in numbers.

Given the negative impact of cognitive biases, it might be tempting to exclusively design explanations that only promote System 2 thinking. This is also risky because it forces users to constantly engage in deliberative thinking, their satisfaction suffers due to higher cognitive friction~\cite{buccinca2020proxy}. We need to strike a balance between System 1 and 2 thinking to appropriately calibrate trust. To do so, we can bridge existing work in non-XAI settings (e.g., balancing System 1 and 2 thinking in clinical decision-making) and translate them to XAI use cases in a contextually relevant manner.

The \textbf{appropriation} of explanatory intent we saw from both groups also has important design implications. Our findings highlight that users will appropriate explanations regardless of careful design. This is not a bad thing. Given dynamic user goals and needs, it is impossible to preemptively exhaust all the interpretations. Our goal then should be to \textit{support, not control} end-users by providing \textit{resources} that mitigate improper or harmful appropriations.

To aid \textit{appropriation-aware} design in XAI, an operationally viable path is to leverage emerging work on Seamful XAI~\cite{ehsan2022seamful} that translates the principles of Seamful Design from Ubiquitous Computing to XAI. Instead of hiding the inherent AI imperfections through seamless design, Seamful XAI turns imperfections into opportunities for explanations while enhancing user agency through appropriation. Simply put, "seams" are mismatches or gaps between what we ideally design the AI to do and how it functions when used in reality. Through a concrete decision process, Seamful XAI uses adversarial thinking like red-teaming to proactively find seams and harness them to improve XAI. For instance, what if the lending officer of an AI-based lending system knew that the AI was trained on North American data but deployed in South Asia (the seam here is a data drift and a context shift)? The knowledge of the mismatch (seam) could help calibrate how they appropriate the AI's output. For instance, they might override the AI's output if it rejects the loan application of a wealthy farmer whose income is non-traditional compared to the income sources in the dataset. In our case, if the AI group were made aware of how little explanatory power Q-values have, it might have prevented them from ascribing diagnostic value to NR's numbers. Last, a key part of designing for appropriation is learning \textit{from} appropriation. Most XAI systems, including ours, are one-way systems that do not allow for user feedback. Feedback loops from users can highlight how, when, and why users are needing to appropriate XAI outputs and ensure the appropriation is appropriate.

\subsection{Broader lessons: Explainable and Responsible AI} \label{sec:broader_lessons}

Our work has broader implications beyond the immediate design of XAI systems, especially around the discourse of responsible and explainable AI. Below, we share three main takeaways—how, despite best intentions, unanticipated negative consequences can emerge from AI-generated explanations \& what we can do about it, how our insights can help re-imagine AI education, and how our insights can reframe how we operationalize AI adoption.   

First, our findings illustrate \textit{how despite best intentions, unsuspecting negative effects of AI-generated explanations can emerge} (e.g., unwarranted faith in numbers in ~\ref{sec:faith_num}). This is an instance of an \textbf{Explainability Pitfall} (EP)---``\textit{unanticipated} negative downstream effects from adding AI explanations that emerge even when there is \textit{no intention} to manipulate anyone''~\cite{ehsan2021explainability}. 
EPs are different from \textit{dark patterns}, where there is an \textit{intent} to deceive the user~\cite{brignull2015dark}. Recall that in our study design, we had no intention of deceiving anyone. 
While dark patterns have been explored in XAI (e.g., ~\cite{chromik2019dark, eiband2018bringing}), \textit{EPs are severely under-explored}. This paper sheds important light on this under-explored space. 

For mitigation strategies, given EPs are often a consequence of \textit{uncritical acceptance}~\cite{ehsan2021explainability}, we can \textit{shift} our explanation design philosophy to emphasize \textit{critical reflection} (as opposed to acceptance) during interpretation of explanations. 
This aligns with Human-centered XAI work that advocates engendering trust via~\textit{reflection}~\cite{ehsan2020human}.
Langer et al.~\cite{langer1978mindlessness} point out that people are likely to accept explanations without conscious attention if no effortful thinking is required from them.
In Kahneman’s dual-process theory~\cite{kahneman2011thinking} terms, this means that if we do not invoke mindful and deliberative (system 2) thinking with explanations, we increase the likelihood of uncritical consumption. 
To trigger mindfulness, Langer et al.~\cite{langer1978mindlessness} recommend designing for “effortful responses” or “thoughtful responding.” 
We can utilize design interventions highlighted in Sec.~\ref{sec:mitigation strategies_6.2} such as Cognitive Forcing Functions and using Seamful XAI strategies.  

Second, our insights call attention to wider challenges with \textit{academic practices of AI learning}. AI students exhibited unwarranted faith in and preference for numbers even when they were not readily comprehensible. In light of the importance of quantitative practices in AI, this bias is not surprising. One effective way to address this would be to critically reflect on the way we educate students in AI. In particular, how do we ensure that students have a more critical eye towards the working and outputs of AI systems? This is where we see the impact of our focus on AI students (vs. fully formed AI practitioners)---by investigating how AI background can lead to the formation of certain heuristics, we have crucial insights on how we might address the issues from an academic training perspective. For instance, there is a need to introduce courses like critical data studies and human-centered data science that can provide much-needed reflective lenses to students to understand their own cognitive biases and the importance of thinking about the user during system design and development. By addressing issues during the formation of their "Data Vision" \cite{Passi2017}, and "Professional Vision" \cite{Goodwin1994} more generally, we can revolutionize how we train the next generations of AI creators and mitigate the creator-consumer gap. If we are asking today’s AI students to become future creators for consumers who think drastically differently from themselves, we need to empower our students with the lenses such that the gap between them and the stakeholders they will serve is less than what it is today. When extrapolating findings for the AI group, we should note that the AI-background population was drawn from students in a very typical introductory AI course, taught from the most widely adopted textbook (by Russell \& Norvig~\cite{russell2010artificial} taught in over 1500 schools worldwide~\cite{aima_russell_norvig_bookadoption}) and using a widely adopted set of instructional assignments.

Third, our work carries broader implications on \textit{reframing AI adoption} through re-examining the relationship between user trust and AI adoption. This re-examination has upstream (e.g., research) and downstream (e..g, industry practices) implications. 
There is an oft-unspoken yet dominant assumption that connects \textit{adoption with acceptance}---where we view AI adoption emerging from user trust, which, in turn, emerges from user acceptance~\cite{ehsan2020human}. 
Acceptance is seen as a core tenet of trust-building (thereby adoption). 
There is often an uncritical push towards trust building without asking a fundamental question: is AI worthy of our trust?
Moreover, \textit{is acceptance the only way to build trust and get adoption?} 
How would we feel in a human relationship if trust hinged solely on accepting everything the other party said?
There are many ways to build trust.
The principles of HCXAI~\cite{ehsan2020human} suggest diverse foundations for trust building such as healthy skepticism from informed users, awareness of variability of around system decisions, and knowledge of what the AI \textit{cannot} do (as opposed to the typical what the AI \textit{can} do). 

Opting to promote reflection in the user can begin the process of defamiliarization from acceptance-first approaches, reframing the discourse around AI adoption. 
Such mindset shifts in AI adoption have cascading societal ripple effects at both upstream (e.g., research) and downstream (e.g., industry practices) levels.
For instance, if an organization embodies a \textit{critically reflective} AI adoption mindset (as opposed to an \textit{acceptance-driven} one), it could mitigate \textit{perverse organizational incentives} such as prioritizing growth and adoption at all costs.
This can reflexively catalyze a change in organizational culture towards responsible and accountable AI governance.
Note that we are not advocating to eliminate building trust via acceptance; rather, we are encouraging reforms that go beyond acceptance.
More importantly, reflection and acceptance are not mutually exclusive and can work in tandem with each other.
Reforming the dialogue around AI adoption and trust building promotes our resilience against detrimental pitfalls like over-estimating (or hyping) AI capabilities, which has cascading societal implications around holding these systems accountable.

Last, when transferring our findings to other XAI scenarios, instead of shooting for blanket generalizability, we should aim for transferability. 
Inherent in transferability is context-sensitivity~\cite{hayes2011relationship,stringer2007action,finfgeld2010generalizability}. 
Real-world XAI settings have different contexts, domains, and application areas ---as such, the transferability of our findings to new domains is subject to context-applicability. Context permitting, especially in terms of user characteristics, the findings are likely to transfer. This is because the heuristics and appropriation strategies identified (in Sec.~\ref{sec:mitigation strategies_6.2}) have broad applicability. For example, our findings can shed light on why data scientists over-trust XAI methods like SHAP that have numerical-based outputs~\cite{kaur2020interpreting}. There are, however, limits to our study design and exhaustiveness of the setup, elaborated in Sec.~\ref{sec:limitations}.

\vspace{-15pt}
\section{Limitations \& Future Work}\label{sec:limitations}
With this study, we have taken a formative step towards understanding how interpretations of AI-generated explanations differ between groups with or without AI background (a consequential user characteristic). 
Given this essential yet first step, the insights from our work should be scoped accordingly. 
A focus on AI students has helped understand how enculturation into AI alters people’s perceptions of AI explanations and makes visible the presence of other kinds of -- often similar -- interpretations within a heterogeneous set of non-AI group participants. However, our study has areas of improvement. First, as shared in Secs.~\ref{sec:screening_and_recruit} \& ~\ref{sec:broader_lessons} while the AI course experienced by the AI group is largely representative of curriculum across other institutions, future research could explore how differences in AI curriculum may impact people's AI background and perception of explanations. Second, in this study, we focused on using an RL-based AI agent to conduct a sequential decision-making task in a controlled environment. Future research could extend to other types of AI agents on other tasks (e.g., classification), especially in a situated sociotechnical context. Third, the quasi-experimental setup (common in behavioral research and neuroscience~\cite{marinescu2018quasi}) does not include random assignment of participants; future work could conduct Randomized Controlled Trials using the same setup to establish causal relationships. Fourth, we used systematic screening to ensure the two groups are measurably different while controlling for factors such as age and education levels (Sec.~\ref{sec:screening_and_recruit}). Despite these efforts, confounds can still exist such as technological familiarity and AI use. Future research can conduct studies 
of user characteristics in isolation and their impact on the interpretation of AI explanations. Fifth, in this initial effort, we scoped our study in the context of three types of AI explanations-- while they are informative, they are not exhaustive of the different types of explanations. 
Future work can explore how users with two or more different user characteristics (e.g., comparison with multiple facets of one's background) or more homogeneous and striated AI backgrounds (e.g., years of AI programming experience) perceive explanations in different ways.
For future iterations, we are inspired by Agre's design philosophy of Critical Technical Practice~\cite{agre1997toward} where \textit{“at least for the foreseeable future, [we] will require a split identity – one foot planted in the craft work of design and the other foot planted in the reflexive work of critique.”}~\cite{agre1997computation}. 
Through this work, we have “planted one foot" in the work of design. Now, we seek to learn from and with the broader HCI and XAI communities as we “plant the other foot” in the self-reflective realm of critique.

\section{Conclusions}
In this paper, we focus on the \textit{who} of XAI 
by investigating how two different groups of \textit{whos}---people with and without a background in AI—perceive different types of AI explanations.
Through a mixed-methods user study, we demonstrate how interpretations and perceptions of AI explanations differ between groups with or without AI background.
Our mixed-methods analysis provides different levels of insight.
\textit{What} people prefer is relatively clear---they prefer natural language-based justificatory rationales.
While the \textit{what} is somewhat straightforward, the \textit{why} and \textit{how} behind their preferences are nuanced. 
Different perceptions (e.g., unanticipated explanatory value) sometimes arise from different forms of appropriation (e.g., diagnosis vs. affirmatory intent).
The same perception (unwarranted faith in numbers) sometimes arises from different types of heuristics (associating numerical vs. incomprehensible reasoning with intelligence)
Finally,  both groups were very similar in their desire to engage with natural-language-based explanations. 

Explainability of AI systems is crucial to instill appropriate user trust and facilitate recourse.
Disparities in AI backgrounds have the potential to exacerbate the challenges arising from the differences between how designers imagine users will appropriate explanations vs. how users actually interpret and use them.
We provided concrete design implications to mitigate the risk of over-reliance on numbers and broader lessons about the need for re-imagining AI education.
By focusing on the who and not just the what of XAI, our work takes a formative step in advancing a pluralistic human-centered XAI discourse to help bridge the creator-consumer gap.

\bibliographystyle{ACM-Reference-Format}
\bibliography{sample-base}

\newpage
\appendix
\section{Appendix}
\subsection{Best practices for data integrity and participant engagement} \label{bestpractices}
Data quality and engagement with user study participants are integral to research. 
Below we share how previous guidelines for fair and equitable work treatment for MTurk workers ~\citep{irani2014critical} helped us decide the payment structure, tips on engagement with participants, design motivations behind the task environment, and how we deployed and reviewed the task catering for a global audience. 
While these insights are transferable to other contexts, most of these practices are geared towards MTurk participants, given one has less control over the platform. 

\begin{enumerate}
    \item \textit{Payment}: our study was not a micro-task that is traditionally deployed in MTurk. 
    As a result, we calibrated our payment to reflect the task duration and expected effort. 
    We tried our best to structure our study based on previous guidelines for fair and equitable work treatment for MTurk workers~\citep{irani2014critical}. 
    We strived to pay equal to or more than a minimum wage [at the time of deployment, the local minimum wage was \$8.5/hour]. 
    We paid \$10 for a task budgeted for 45 minutes, making the hourly pay \$13.3. However, almost all participants took 30 mins to complete the task on average, making the effective hourly rate around \$20/hour.
    
    As a policy, we disbursed payments within 48 hours of task completion. 
    This robust turnaround time helped our reputation on Turkopticon, a forum for MTurk workers to engage in peer-to-peer assistance on job information and hold employers accountable for fair treatment~\citep{irani2014critical}. 
    Moreover, one researcher regularly engaged with workers on Turkopticon, answering questions and returning compliments. 
    This engagement built rapport throughout the study. As a platform, TurkPrime allows internal messages between MTurk workers and employers by using a proxy userID and protecting the privacy of the worker. 
    This feature allowed participants to communicate if they had internet issues or were running out of time. 
    In such cases, we sent them a one-time link for completion. 
    The same applied for participants in the AI background group who were not bound by AMT rules.
    
    Every payment of a HIT had a thank you message attached to it. Every rejection had custom justifications backed by evidence. 
    The research team created message templates based on major issues, which allowed for a quick turnaround time even with a custom message. 
    For participants who failed to do the task despite best efforts, we paid them for their time even if we could not use their data. 
    This equitable policy also made our HITs one of the most sought-after in the marketplace. 
    \item \textit{Task environment and setup:}
    
    \begin{enumerate}
        \item \textit{Task orientation}: Pilot testing showed that participants preferred a multi-modal (e.g., video) orientation compared to a textual description of it. 
        Therefore, we provided both modalities.
        \item \textit{Task engagement}: Participant had to successfully pass attention checks and/or explicitly acknowledge they understood the instructions in a given module. 
        On the backend, we had timers to evaluate if a participant spent a reasonable amount of time on a particular section. 
        For instance, if the video was 2 minutes long and a participant clicked through that section in 30 seconds, it triggered a review of their responses. 
        These steps augment the quality of the experimental data. 
        Moreover, for tasks that had to be rejected, these metrics also served as justificatory evidence.
        \item \textit{Design of the robots}: To mitigate effects of preconceived notions, we did not use any descriptive names for the robots; instead, we introduced the robots as ``Robot A'' [= the Rationale-Generation robot], ``Robot B'' [= the Action-Declaring robot], and ``Robot C'' [= the Numerical robot] (details on each of their attributes are below/above). 
        To reduce any preferential treatment of robots based on their appearances, we need to standardize their appearances without sacrificing their distinctness. That is, the robots needed to look similar, but not to the point of indistinguishability. 
        This insight came from pilot testing which indicated that if we made the robots identical in appearance but different in color, the cognitive load for recall was too high. 
        Therefore, we iterated and struck a balance where all robots had wheels with a ``car-like'' structure~(see Fig. \ref{fig:3robots}) while they differed in color and shape.
    \end{enumerate}
    
    \item \textit{Deployment and Review}: Across both groups, we manually reviewed every response in the survey, especially the qualitative justifications provided by the participants. 
    We deployed 10-15 tasks per day to allow for manual reviews. 
    To facilitate outreach of our task to all time zones, using an automated scheduling system, we released 3 tasks every 3 hours over a 24-hour cycle. 
    This improved the potential for global participation in our task. 
    
    Spamming is a serious issue when it comes to survey data. 
    Here, the qualitative responses served a secondary screening purpose. 
    Participants with good-faith efforts always had reasonable qualitative justifications. 
    Those who had spamming intentions shared non-sensical and even comical qualitative responses; e.g., Movie titles and plots, snippets from Wikipedia, etc.
\end{enumerate}
While all of these steps required considerable time, and effort, they paid off in the high data quality we received for a task lasting 30 minutes on average.

\newpage

\subsection{Participant Screening} \label{partcipant screener}

\paragraph{\textbf{Screening criteria, group makeup, and establishing group differences:}}
\hfill

To ensure that the two groups were \textit{measurably different} along the dimension of our investigation---AI background---we performed additional \textit{screening} using a questionnaire with three components---(1)~a knowledge test to get a baseline understanding of programming and AI competency. This test was collaboratively developed with the course's teaching staff to calibrate the content relevancy and question difficulty.
(2)~ Self-reported knowledge levels in (a)~computer programming and (b)~AI using two 5-point Likert-scales.
(3)~ confirmation of whether they have ever taken an AI class.

To get empirically ground the cut-off points, we piloted the screener with 10 participants from each group to get a baseline understanding of the scores and completion time. 
For the \textit{AI} group, a score 4 or more on the knowledge test along with self-reports of having ``Moderate knowledge'' or more $[>=4]$ in programming and some knowledge or more $[>=3]$ in AI were required.
For the \textit{non-AI} group, self-report of having ``No knowledge'' $[=1]$ in both programming and AI, along with no prior AI classes, were required. Using these criteria we formed the two groups. 

The \textit{AI background group} consisted of 96 adult students taking an AI class. On average, the task duration was $31.1$ minutes.
Participants received US $\$10$ for their time (\$20/hr rate).
$39\%$ of the participants self-identified as females while the rest identified as males.
Participants reported an average education level of $5.25$ (5= “Associate’s degree”, 6= “Bachelor’s Degree”). 
By design, all of them currently reside in the US. 
On the \textit{screening criteria}, the AI students scored an average of 4.73 (out of 5) [$SD= 0.45$] on the \textit{knowledge test}, self-reported ``moderate knowledge'' on \textit{programming}~($M=4.53, SD=0.52$) [4= “Moderate Knowledge”, 5= “A lot of knowledge”]  and ``some knowledge'' on\textit{AI}~($M=3.74, SD= 0.49$) [3= “Some knowledge”, 4= “Moderate Knowledge” ].

The {\em Non-AI background group with MTurk participants} consisted of 83 adults, who were recruited from Amazon Mechanical Turk (AMT) through a management service called TurkPrime~\cite{litman2017turkprime}. 
On average, the task duration was $29.8$ minutes. 
Participants received US $\$10$ for their time (\$20/hr).
$46\%$ of the participants self-identified as females while the rest identified as males. 
Participants reported an average education level of $4.8$ (4= “Vocational Training”, 5= Associate’s degree).
We screened for participants who reside in the US.
On the \textit{screening criteria}, MTurk participants scored an average of $0.91$ (out of 5) [$SD=0.32$] on the \textit{knowledge test}. By design, for \textit{programming} and \textit{AI}, we screened for people who reported as having ``No knowledge'' [=1] as well as \textit{never taking} an AI class.

To establish that these \textit{two groups are measurably different}, we performed statistical tests. For the knowledge test, the two groups were significantly different based on a two-sample Mann-Whitney U-test (even after Bonferroni correction, $p <  2.2\times~10^{-16}$). 
Since the non-AI group systematically had only people with “No knowledge”~[=1] in programming or AI, we performed one-sample Mann Whitney U-tests on the AI-group to compare its means against~1[=``No knowledge'']. 
Even after Bonferroni correction, we found strong evidence that the two groups are different ($p <  2.2\times 10^{-16}$, for both programming and AI scores). These results indicate that our screening criteria have successfully established two groups that are measurably different in terms of their AI background.

The following aspects of the selection criteria facilitated formation of the two user groups:

\paragraph{\textbf{The AI knowledge questionnaire}:} 

We developed the knowledge questionnaire iteratively using a participatory process involving Teaching Assistants for the class along with Graduate students familiar with the area. 
There are five (5) multiple-choice questions in total equalling 5 points. 
The first two questions are programming questions: the first is a question asking for the output of a simple print statement, and the second is one that asks for the output of a for-loop. 
The remaining three covered concepts in AI such as Markov Decision Processes, Reinforcement Learning, and Unsupervised Learning. 
By the time of deployment, the AI students had already gone through lectures covering the AI topics in the questionnaire. 
All the questions are inspired by or directly taken from past exam questions on various topics.
For further details, please refer to section \ref{screenerQ} for the AI knowledge questionnaire. 

To calibrate the relative difficulty of the knowledge test, we used a collaborative and iterative process until a consensus between the researchers and the teaching staff was reached. 
We expected that most students with satisfactory prerequisites (that contain fundamentals of programming) and current knowledge from the class should at least get 4 out of the 5 correct. 
This calibration appears to have been a reasonable one since all students naturally passed these thresholds. 
On the other hand, in order to be assigned to the non-AI background group, participants had to score less than or equal to 1 (out of 5). 
We expected that some participants, without any AI background, might be able to guess the output of the “print” statement question. 
However, it was unlikely that someone without a basic programming understanding would be able to answer the “for-loop” output question. 
Therefore, if someone correctly answers more than one, their AI knowledge background is not the type we need for members of the non-AI background group. 

\paragraph{\textbf{AI background measurement:}} 
The knowledge test was followed by two 5-point Likert-scale questions measuring the AI background for computer programming and AI concepts. 
The range of self-reported knowledge goes from "No knowledge"$[=1]$ to "A lot of knowledge" $[=5]$. 
Each level of knowledge has a sentence clarifying the meaning behind the label.
For illustrative purposes, here is an example from the AI scale: "No knowledge: I might be \textit{aware} of AI, but have \textit{no knowledge} about it." 
Both scales had similar construction and wording. 
For further details, please refer to section \ref{screenerQ} for the AI background knowledge Likert-scales. 

\paragraph{AI class:} Finally, participants answer if they have ever taken any classes on Artificial Intelligence. 

\subsubsection{\textbf{Screening questionnaire}} \label{screenerQ}
\hfill\\
Here we share the survey instruments used to screen participants.
\pseudosection {Knowledge test questionnaire}
\begin{enumerate}
    \item What would be the output of the following python program?
    \begin{verbatim}
    name = "Peter"
    print("Hello " + name)
    \end{verbatim}
    \begin{enumerate}
        \item Peter
        \item Hello Peter
        \item Hello + Peter
        \item "Hello" + name
    \end{enumerate}
    \item What would be the output of the following python program?
    \begin{verbatim}
    numbers = [2, 4]
    for i in range(len(numbers)):
	        print(numbers[i] + i)
    \end{verbatim}
    \begin{enumerate}
        \item 2
        
        5
        \item 2
        
        5
        
        8
        \item2
        
        4
        \item2
        
        4
        
        10
    \end{enumerate}
    \item Which of the following is an unsupervised learning task?
    \begin{enumerate}
        \item Distinguishing pictures containing cats from pictures not containing cats
        \item Flagging text messages as appropriate or inappropriate
        \item Divide data points into different clusters without any labels available
        \item Predict the value of a house after training on a dataset with house features and values
    \end{enumerate}
    \item What is the general goal of reinforcement learning?
    \begin{enumerate}
        \item Maximize potential or expected punishment
        \item Maximize potential or expected reward
        \item Get to the goal as soon as possible
        \item Avoid the most obstacles in any given state 
    \end{enumerate}
    \item In MDPs, the Markov assumption is that:
    \begin{enumerate}
        \item The current state is independent of all other states
        \item The current state depends only on the history of previous states and actions
        \item The current state depends on the full sequence of states and actions (past and future)
        \item The current state only depends on the immediate previous state and action
    \end{enumerate}
\end{enumerate}
\subsubsection*{Computer Programming Background Knowledge}
\hfill\\
When it comes to computer programming or coding, I believe I have
\begin{enumerate}
    \item No knowledge: I might be \textit{aware} of computer programs, but have \textit{never} coded before
    \item A little knowledge: I know \textit{basic} concepts in programming, but have \textit{never applied} it
    \item Some knowledge: I have \textit{applied} programming concepts by coding at least \textit{once} before 
    \item Moderate knowledge: I \textit{apply} programming concepts \textit{somewhat frequently} for my work, class, or leisure
    \item A lot of knowledge: I \textit{apply} programming concepts \textit{very frequently} or create cutting edge software 
\end{enumerate}

\subsubsection*{AI Background Knowledge} 
\hfill\\
When it comes to Artificial Intelligence (AI), I believe I have
\begin{enumerate}
    \item No knowledge: I might be \textit{aware} of AI, but have \textit{no knowledge} about it
    \item A little knowledge: I know \textit{basic} concepts in AI, but have \textit{never applied} it
    \item Some knowledge: I have \textit{applied} AI concepts by coding at least \textit{once} before 
    \item Moderate knowledge: I \textit{apply} AI concepts \textit{somewhat frequently} for my work, class, or leisure
    \item A lot of knowledge: I \textit{apply} AI concepts \textit{very frequently} or create cutting edge software
\end{enumerate}

\subsubsection*{AI class}
\hfill\\
Have you ever taken or are currently taking any classes on Artificial Intelligence?
\begin{itemize}
    \item Yes
    \item No
\end{itemize}

\newpage

\newpage
\subsection{OLR Summary Tables} \label{OLR tables}

\begin{table*}[tbh]
\caption{\label{tab:R1} Summary of OLR with Ranking as Response and Robot Type as Predictor\label{tab:Overall_Robot_Ranking_RG baseline}
}
\centering
\begin{tabular}[t]{lccccc}
\toprule
  & Value & Std. Error & $t$- value & $p$- value & Odds Ratio\\
\midrule
\textbf{AD Robot} & -1.711 & 0.104 & -16.473 & \textbf{< 0.001} & \textbf{0.181}\\
\textbf{NR\_Robot} & -2.691 & 0.115 & -23.393 & \textbf{< 0.001} & \textbf{0.068}\\
1|2 & -2.366 & 0.093 & -25.567 & < 0.001 & 0.094\\
2|3 & -0.598 & 0.077 & -7.799 & < 0.001 & 0.5501\\
\bottomrule
\multicolumn{6}{l}{\textit{Note: } RG\_Robot is the reference level.}\\
\end{tabular}
\end{table*}

\begin{table*}[tbh]
    \caption{\label{tab:kable2}OLR Summary with Robot Type\label{tab:Overall_Robot_Ranking_AD baseline}}
    \centering
    \begin{tabular}[t]{lccccc}
    \toprule
      & Value & Std. Error & t value & p value & Odds Ratio\\
    \midrule
    Rationale-Generation Robot & 1.7105 & 0.1038 & 16.4728 & 0 & 5.5319\\
    Numerical-Reasoning Robot & -0.9807 & 0.1001 & -9.7980 & 0 & 0.3751\\
    1|2 & -0.6550 & 0.0695 & -9.4282 & 0 & 0.5195\\
    2|3 & 1.1129 & 0.0734 & 15.1592 & 0 & 3.0433\\
    \bottomrule
    \multicolumn{6}{l}{\textit{Note: } Action-Declaring Robot is the reference level.}\\
    \end{tabular}
\end{table*}

\begin{table*}[tbh]
\caption{\label{tab:kable2}OLR Summary - Ref. Levels: Rationale-Generation Robot and  AI Group\label{tab:ANon-Turk}}
\centering
\begin{tabular}[t]{lccccc}
\toprule
  & Value & Std. Error & t value & p value & Odds Ratio\\
\midrule
Action-Declaring Robot & -2.0246 & 0.1302 & -15.5462 & 0.0000 & 0.1320\\
Numerical-Reasoning Robot & -2.5072 & 0.1391 & -18.0221 & 0.0000 & 0.0815\\
Non-AI Group & -0.1563 & 0.1582 & -0.9879 & 0.3232 & 0.8553\\
Action-Declaring Robot:Non-AI Group & 0.8422 & 0.2093 & 4.0243 & 0.0001 & 2.3214\\
Numerical-Reasoning Robot:Non-AI Group & -0.6088 & 0.2264 & -2.6892 & 0.0072 & 0.5440\\
\addlinespace
1|2 & -2.4515 & 0.1104 & -22.2009 & 0.0000 & 0.0862\\
2|3 & -0.6507 & 0.0965 & -6.7413 & 0.0000 & 0.5217\\
\bottomrule
\end{tabular}
\end{table*}

\begin{table*}[tbh]

\caption{\label{tab:kable2}OLR Summary - Ref. Levels: Rationale-Generation Robot and  Non-AI Group\label{tab:ATurk}}
\centering
\begin{tabular}[t]{lccccc}
\toprule
  & Value & Std. Error & t value & p value & Odds Ratio\\
\midrule
Action-Declaring Robot & -1.1824 & 0.1674 & -7.0614 & 0.0000 & 0.3065\\
Numerical-Reasoning Robot & -3.1160 & 0.1893 & -16.4636 & 0.0000 & 0.0443\\
AI Group & 0.1564 & 0.1582 & 0.9881 & 0.3231 & 1.1692\\
Action-Declaring Robot:AI Group & -0.8422 & 0.2093 & -4.0245 & 0.0001 & 0.4308\\
Numerical-Reasoning Robot:AI Group & 0.6087 & 0.2264 & 2.6889 & 0.0072 & 1.8381\\
\addlinespace
1|2 & -2.2952 & 0.1362 & -16.8540 & 0.0000 & 0.1007\\
2|3 & -0.4944 & 0.1258 & -3.9305 & 0.0001 & 0.6099\\
\bottomrule
\end{tabular}
\end{table*}

\begin{table*}[tbh]

\caption{\label{tab:kable2}OLR Summary - Ref. Levels: Action-Declaring Robot and  AI Group\label{tab:BNon-Turk}}
\centering
\begin{tabular}[t]{lccccc}
\toprule
  & Value & Std. Error & t value & p value & Odds Ratio\\
\midrule
Rationale-Generation Robot & 2.0246 & 0.1302 & 15.5462 & 0e+00 & 7.5732\\
Numerical-Reasoning Robot & -0.4826 & 0.1224 & -3.9421 & 1e-04 & 0.6172\\
Non-AI Group & 0.6859 & 0.1368 & 5.0127 & 0e+00 & 1.9855\\
Rationale-Generation Robot:Non-AI Group & -0.8422 & 0.2093 & -4.0243 & 1e-04 & 0.4308\\
Numerical-Reasoning Robot:Non-AI Group & -1.4510 & 0.2125 & -6.8278 & 0e+00 & 0.2343\\
\addlinespace
1|2 & -0.4269 & 0.0836 & -5.1091 & 0e+00 & 0.6526\\
2|3 & 1.3739 & 0.0904 & 15.2045 & 0e+00 & 3.9507\\
\bottomrule
\end{tabular}
\end{table*}

\begin{table*}[tbh]

\caption{\label{tab:kable2}OLR Summary - Ref. Levels: Action-Declaring Robot and  Non-AI Group\label{tab:BTurk}}
\centering
\begin{tabular}[t]{lccccc}
\toprule
  & Value & Std. Error & t value & p value & Odds Ratio\\
\midrule
Rationale-Generation Robot & 1.1824 & 0.1674 & 7.0613 & 0e+00 & 3.2622\\
Numerical-Reasoning Robot & -1.9336 & 0.1750 & -11.0468 & 0e+00 & 0.1446\\
AI Group & -0.6859 & 0.1368 & -5.0127 & 0e+00 & 0.5037\\
Rationale-Generation Robot:AI Group & 0.8422 & 0.2093 & 4.0245 & 1e-04 & 2.3215\\
Numerical-Reasoning Robot:AI Group & 1.4510 & 0.2125 & 6.8278 & 0e+00 & 4.2672\\
\addlinespace
1|2 & -1.1127 & 0.1147 & -9.6973 & 0e+00 & 0.3287\\
2|3 & 0.6880 & 0.1123 & 6.1249 & 0e+00 & 1.9898\\
\bottomrule
\end{tabular}
\end{table*}

\begin{table*}[tbh]

\caption{\label{tab:kable2}OLR Summary - Ref. Levels: Numerical-Reasoning Robot and  AI Group\label{tab:CNon-Turk}}
\centering
\begin{tabular}[t]{lccccc}
\toprule
  & Value & Std. Error & t value & p value & Odds Ratio\\
\midrule
Rationale-Generation Robot & 2.5072 & 0.1391 & 18.0222 & 0.0000 & 12.2708\\
Action-Declaring Robot & 0.4826 & 0.1224 & 3.9422 & 0.0001 & 1.6203\\
Non-AI Group & -0.7651 & 0.1620 & -4.7240 & 0.0000 & 0.4653\\
Rationale-Generation Robot:Non-AI Group & 0.6088 & 0.2264 & 2.6891 & 0.0072 & 1.8382\\
Action-Declaring Robot:Non-AI Group & 1.4510 & 0.2125 & 6.8278 & 0.0000 & 4.2672\\
\addlinespace
1|2 & 0.0557 & 0.0923 & 0.6037 & 0.5460 & 1.0573\\
2|3 & 1.8565 & 0.1033 & 17.9798 & 0.0000 & 6.4012\\
\bottomrule
\end{tabular}
\end{table*}

\begin{table*}[tbh]

\caption{\label{tab:kable2}OLR Summary - Ref. Levels: Numerical-Reasoning Robot and  Non-AI Group\label{tab:CTurk}}
\centering
\begin{tabular}[t]{lccccc}
\toprule
  & Value & Std. Error & t value & p value & Odds Ratio\\
\midrule
Rationale-Generation Robot & 3.1160 & 0.1893 & 16.4637 & 0.0000 & 22.5559\\
Action-Declaring Robot & 1.9336 & 0.1750 & 11.0468 & 0.0000 & 6.9141\\
AI Group & 0.7651 & 0.1620 & 4.7240 & 0.0000 & 2.1492\\
Rationale-Generation Robot:AI Group & -0.6088 & 0.2264 & -2.6891 & 0.0072 & 0.5440\\
Action-Declaring Robot:AI Group & -1.4510 & 0.2125 & -6.8279 & 0.0000 & 0.2343\\
\addlinespace
1|2 & 0.8208 & 0.1336 & 6.1447 & 0.0000 & 2.2724\\
2|3 & 2.6216 & 0.1444 & 18.1523 & 0.0000 & 13.7575\\
\bottomrule
\end{tabular}
\end{table*}

\begin{table*}[tbh]

\caption{\label{tab:kable3}OLR Summary -  Confidence  - Ref. Levels: Type Rationale-Generation Robot and Group AI Group\label{tab:Confidence_ANon-Turk}}
\centering
\begin{tabular}[t]{lccccc}
\toprule
  & Value & Std. Error & t value & p value & Odds Ratio\\
\midrule
Action-Declaring Robot & -1.3753 & 0.2740 & -5.0185 & 0.0000 & 0.2528\\
Numerical-Reasoning Robot & -1.2607 & 0.2815 & -4.4783 & 0.0000 & 0.2835\\
Non-AI Group & -0.0290 & 0.3297 & -0.0879 & 0.9299 & 0.9714\\
Action-Declaring Robot:Non-AI Group & 0.7232 & 0.4545 & 1.5913 & 0.1115 & 2.0610\\
Numerical-Reasoning Robot:Non-AI Group & -0.6906 & 0.4668 & -1.4795 & 0.1390 & 0.5013\\
\addlinespace
1|2 & -1.6705 & 0.2164 & -7.7194 & 0.0000 & 0.1882\\
2|3 & -0.1251 & 0.1996 & -0.6268 & 0.5308 & 0.8824\\
\bottomrule
\end{tabular}
\end{table*}

\begin{table*}[tbh]

\caption{\label{tab:kable3}OLR Summary -  Confidence  - Ref. Levels: Type Rationale-Generation Robot and Group Non-AI Group\label{tab:Confidence_ATurk}}
\centering
\begin{tabular}[t]{lccccc}
\toprule
  & Value & Std. Error & t value & p value & Odds Ratio\\
\midrule
Action-Declaring Robot & -0.6521 & 0.3654 & -1.7845 & 0.0743 & 0.5210\\
Numerical-Reasoning Robot & -1.9513 & 0.3800 & -5.1343 & 0.0000 & 0.1421\\
AI Group & 0.0290 & 0.3297 & 0.0879 & 0.9299 & 1.0294\\
Action-Declaring Robot:AI Group & -0.7232 & 0.4545 & -1.5913 & 0.1115 & 0.4852\\
Numerical-Reasoning Robot:AI Group & 0.6906 & 0.4668 & 1.4795 & 0.1390 & 1.9949\\
\addlinespace
1|2 & -1.6415 & 0.2777 & -5.9112 & 0.0000 & 0.1937\\
2|3 & -0.0961 & 0.2651 & -0.3626 & 0.7169 & 0.9083\\
\bottomrule
\end{tabular}
\end{table*}

\begin{table*}[tbh]

\caption{\label{tab:kable3}OLR Summary -  Confidence  - Ref. Levels: Type Action-Declaring Robot and Group AI Group\label{tab:Confidence_BNon-Turk}}
\centering
\begin{tabular}[t]{lccccc}
\toprule
  & Value & Std. Error & t value & p value & Odds Ratio\\
\midrule
Rationale-Generation Robot & 1.3753 & 0.2740 & 5.0185 & 0.0000 & 3.9561\\
Numerical-Reasoning Robot & 0.1146 & 0.2678 & 0.4279 & 0.6687 & 1.1214\\
Non-AI Group & 0.6942 & 0.3127 & 2.2203 & 0.0264 & 2.0021\\
Rationale-Generation Robot:Non-AI Group & -0.7232 & 0.4545 & -1.5913 & 0.1115 & 0.4852\\
Numerical-Reasoning Robot:Non-AI Group & -1.4138 & 0.4561 & -3.0995 & 0.0019 & 0.2432\\
\addlinespace
1|2 & -0.2952 & 0.1873 & -1.5759 & 0.1150 & 0.7444\\
2|3 & 1.2501 & 0.1979 & 6.3175 & 0.0000 & 3.4908\\
\bottomrule
\end{tabular}
\end{table*}

\begin{table*}[tbh]

\caption{\label{tab:kable3}OLR Summary -  Confidence  - Ref. Levels: Type Action-Declaring Robot and Group Non-AI Group\label{tab:Confidence_BTurk}}
\centering
\begin{tabular}[t]{lccccc}
\toprule
  & Value & Std. Error & t value & p value & Odds Ratio\\
\midrule
Rationale-Generation Robot & 0.6521 & 0.3654 & 1.7845 & 0.0743 & 1.9196\\
Numerical-Reasoning Robot & -1.2992 & 0.3688 & -3.5230 & 0.0004 & 0.2727\\
AI Group & -0.6942 & 0.3127 & -2.2203 & 0.0264 & 0.4995\\
Rationale-Generation Robot:AI Group & 0.7232 & 0.4545 & 1.5913 & 0.1115 & 2.0609\\
Numerical-Reasoning Robot:AI Group & 1.4138 & 0.4561 & 3.0995 & 0.0019 & 4.1115\\
\addlinespace
1|2 & -0.9894 & 0.2603 & -3.8015 & 0.0001 & 0.3718\\
2|3 & 0.5560 & 0.2566 & 2.1668 & 0.0303 & 1.7436\\
\bottomrule
\end{tabular}
\end{table*}

\begin{table*}[tbh]

\caption{\label{tab:kable3}OLR Summary -  Confidence  - Ref. Levels: Type Numerical-Reasoning Robot and Group AI Group\label{tab:Confidence_CNon-Turk}}
\centering
\begin{tabular}[t]{lccccc}
\toprule
  & Value & Std. Error & t value & p value & Odds Ratio\\
\midrule
Rationale-Generation Robot & 1.2607 & 0.2815 & 4.4783 & 0.0000 & 3.5278\\
Action-Declaring Robot & -0.1146 & 0.2678 & -0.4279 & 0.6687 & 0.8918\\
Non-AI Group & -0.7196 & 0.3304 & -2.1778 & 0.0294 & 0.4870\\
Rationale-Generation Robot:Non-AI Group & 0.6906 & 0.4668 & 1.4795 & 0.1390 & 1.9949\\
Action-Declaring Robot:Non-AI Group & 1.4138 & 0.4561 & 3.0995 & 0.0019 & 4.1115\\
\addlinespace
1|2 & -0.4098 & 0.1999 & -2.0501 & 0.0404 & 0.6638\\
2|3 & 1.1356 & 0.2076 & 5.4701 & 0.0000 & 3.1130\\
\bottomrule
\end{tabular}
\end{table*}

\begin{table*}[tbh]

\caption{\label{tab:kable3}OLR Summary -  Confidence  - Ref. Levels: Type Numerical-Reasoning Robot and Group Non-AI Group\label{tab:Confidence_CTurk}}
\centering
\begin{tabular}[t]{lccccc}
\toprule
  & Value & Std. Error & t value & p value & Odds Ratio\\
\midrule
Rationale-Generation Robot & 1.9513 & 0.3800 & 5.1343 & 0.0000 & 7.0377\\
Action-Declaring Robot & 1.2992 & 0.3688 & 3.5230 & 0.0004 & 3.6664\\
AI Group & 0.7196 & 0.3304 & 2.1778 & 0.0294 & 2.0536\\
Rationale-Generation Robot:AI Group & -0.6906 & 0.4668 & -1.4795 & 0.1390 & 0.5013\\
Action-Declaring Robot:AI Group & -1.4138 & 0.4561 & -3.0995 & 0.0019 & 0.2432\\
\addlinespace
1|2 & 0.3098 & 0.2666 & 1.1620 & 0.2452 & 1.3631\\
2|3 & 1.8552 & 0.2812 & 6.5966 & 0.0000 & 6.3927\\
\bottomrule
\end{tabular}
\end{table*}

\begin{table*}[tbh]

\caption{\label{tab:kable3}OLR Summary -  Friendliness  - Ref. Levels: Type Rationale-Generation Robot and Group AI Group\label{tab:Friendliness_ANon-Turk}}
\centering
\begin{tabular}[t]{lccccc}
\toprule
  & Value & Std. Error & t value & p value & Odds Ratio\\
\midrule
Action-Declaring Robot & -5.8425 & 0.6241 & -9.3621 & 0.0000 & 0.0029\\
Numerical-Reasoning Robot & -9.1613 & 0.6886 & -13.3037 & 0.0000 & 0.0001\\
Non-AI Group & -1.2732 & 0.6531 & -1.9494 & 0.0512 & 0.2799\\
Action-Declaring Robot:Non-AI Group & 2.3121 & 0.7844 & 2.9477 & 0.0032 & 10.0953\\
Numerical-Reasoning Robot:Non-AI Group & 0.4423 & 0.8811 & 0.5021 & 0.6156 & 1.5564\\
\addlinespace
1|2 & -7.4844 & 0.6292 & -11.8956 & 0.0000 & 0.0006\\
2|3 & -3.1140 & 0.5110 & -6.0945 & 0.0000 & 0.0444\\
\bottomrule
\end{tabular}
\end{table*}

\begin{table*}[tbh]

\caption{\label{tab:kable3}OLR Summary -  Friendliness  - Ref. Levels: Type Rationale-Generation Robot and Group Non-AI Group\label{tab:Friendliness_ATurk}}
\centering
\begin{tabular}[t]{lccccc}
\toprule
  & Value & Std. Error & t value & p value & Odds Ratio\\
\midrule
Action-Declaring Robot & -3.5301 & 0.5314 & -6.6431 & 0.0000 & 0.0293\\
Numerical-Reasoning Robot & -8.7179 & 0.7549 & -11.5482 & 0.0000 & 0.0002\\
AI Group & 1.2734 & 0.6531 & 1.9497 & 0.0512 & 3.5730\\
Action-Declaring Robot:AI Group & -2.3128 & 0.7844 & -2.9485 & 0.0032 & 0.0990\\
Numerical-Reasoning Robot:AI Group & -0.4435 & 0.8809 & -0.5035 & 0.6146 & 0.6418\\
\addlinespace
1|2 & -6.2112 & 0.5476 & -11.3434 & 0.0000 & 0.0020\\
2|3 & -1.8407 & 0.4068 & -4.5247 & 0.0000 & 0.1587\\
\bottomrule
\end{tabular}
\end{table*}

\begin{table*}[tbh]

\caption{\label{tab:kable3}OLR Summary -  Friendliness  - Ref. Levels: Type Action-Declaring Robot and Group AI Group\label{tab:Friendliness_BNon-Turk}}
\centering
\begin{tabular}[t]{lccccc}
\toprule
  & Value & Std. Error & t value & p value & Odds Ratio\\
\midrule
Rationale-Generation Robot & 5.8429 & 0.6241 & 9.3622 & 0.0000 & 344.7825\\
Numerical-Reasoning Robot & -3.3185 & 0.3827 & -8.6712 & 0.0000 & 0.0362\\
Non-AI Group & 1.0394 & 0.4342 & 2.3940 & 0.0167 & 2.8274\\
Rationale-Generation Robot:Non-AI Group & -2.3129 & 0.7844 & -2.9486 & 0.0032 & 0.0990\\
Numerical-Reasoning Robot:Non-AI Group & -1.8692 & 0.7336 & -2.5481 & 0.0108 & 0.1542\\
\addlinespace
1|2 & -1.6417 & 0.2599 & -6.3158 & 0.0000 & 0.1936\\
2|3 & 2.7288 & 0.3585 & 7.6120 & 0.0000 & 15.3144\\
\bottomrule
\end{tabular}
\end{table*}

\begin{table*}[tbh]

\caption{\label{tab:kable3}OLR Summary -  Friendliness  - Ref. Levels: Type Action-Declaring Robot and Group Non-AI Group\label{tab:Friendliness_BTurk}}
\centering
\begin{tabular}[t]{lccccc}
\toprule
  & Value & Std. Error & t value & p value & Odds Ratio\\
\midrule
Rationale-Generation Robot & 3.5301 & 0.5314 & 6.6431 & 0.0000 & 34.1267\\
Numerical-Reasoning Robot & -5.1877 & 0.6660 & -7.7898 & 0.0000 & 0.0056\\
AI Group & -1.0394 & 0.4341 & -2.3941 & 0.0167 & 0.3537\\
Rationale-Generation Robot:AI Group & 2.3125 & 0.7844 & 2.9483 & 0.0032 & 10.1000\\
Numerical-Reasoning Robot:AI Group & 1.8692 & 0.7336 & 2.5481 & 0.0108 & 6.4832\\
\addlinespace
1|2 & -2.6811 & 0.4164 & -6.4392 & 0.0000 & 0.0685\\
2|3 & 1.6894 & 0.3421 & 4.9375 & 0.0000 & 5.4160\\
\bottomrule
\end{tabular}
\end{table*}

\begin{table*}[tbh]

\caption{\label{tab:kable3}OLR Summary -  Friendliness  - Ref. Levels: Type Numerical-Reasoning Robot and Group AI Group\label{tab:Friendliness_CNon-Turk}}
\centering
\begin{tabular}[t]{lccccc}
\toprule
  & Value & Std. Error & t value & p value & Odds Ratio\\
\midrule
Rationale-Generation Robot & 9.1614 & 0.6887 & 13.3033 & 0.0000 & 9522.2302\\
Action-Declaring Robot & 3.3185 & 0.3827 & 8.6712 & 0.0000 & 27.6181\\
Non-AI Group & -0.8299 & 0.5912 & -1.4038 & 0.1604 & 0.4361\\
Rationale-Generation Robot:Non-AI Group & -0.4436 & 0.8809 & -0.5036 & 0.6146 & 0.6417\\
Action-Declaring Robot:Non-AI Group & 1.8693 & 0.7336 & 2.5481 & 0.0108 & 6.4835\\
\addlinespace
1|2 & 1.6768 & 0.2812 & 5.9620 & 0.0000 & 5.3482\\
2|3 & 6.0473 & 0.4618 & 13.0936 & 0.0000 & 422.9598\\
\bottomrule
\end{tabular}
\end{table*}

\begin{table*}[tbh]

\caption{\label{tab:kable3}OLR Summary -  Friendliness  - Ref. Levels: Type Numerical-Reasoning Robot and Group Non-AI Group\label{tab:Friendliness_CTurk}}
\centering
\begin{tabular}[t]{lccccc}
\toprule
  & Value & Std. Error & t value & p value & Odds Ratio\\
\midrule
Rationale-Generation Robot & 8.7179 & 0.7549 & 11.5480 & 0.0000 & 6111.4635\\
Action-Declaring Robot & 5.1879 & 0.6660 & 7.7897 & 0.0000 & 179.0894\\
AI Group & 0.8301 & 0.5912 & 1.4041 & 0.1603 & 2.2934\\
Rationale-Generation Robot:AI Group & 0.4434 & 0.8809 & 0.5033 & 0.6148 & 1.5579\\
Action-Declaring Robot:AI Group & -1.8695 & 0.7336 & -2.5483 & 0.0108 & 0.1542\\
\addlinespace
1|2 & 2.5068 & 0.5200 & 4.8208 & 0.0000 & 12.2655\\
2|3 & 6.8773 & 0.6364 & 10.8059 & 0.0000 & 970.0040\\
\bottomrule
\end{tabular}
\end{table*}

\begin{table*}[tbh]

\caption{\label{tab:kable3}OLR Summary -  Intelligence  - Ref. Levels: Type Rationale-Generation Robot and Group AI Group\label{tab:Intelligence_ANon-Turk}}
\centering
\begin{tabular}[t]{lccccc}
\toprule
  & Value & Std. Error & t value & p value & Odds Ratio\\
\midrule
Action-Declaring Robot & -1.8632 & 0.2799 & -6.6555 & 0.0000 & 0.1552\\
Numerical-Reasoning Robot & -0.9546 & 0.2801 & -3.4088 & 0.0007 & 0.3850\\
Non-AI Group & -0.3321 & 0.3304 & -1.0053 & 0.3147 & 0.7174\\
Action-Declaring Robot:Non-AI Group & 1.0047 & 0.4542 & 2.2120 & 0.0270 & 2.7310\\
Numerical-Reasoning Robot:Non-AI Group & -0.1047 & 0.4626 & -0.2262 & 0.8210 & 0.9006\\
\addlinespace
1|2 & -1.7485 & 0.2186 & -7.9976 & 0.0000 & 0.1740\\
2|3 & -0.2121 & 0.2011 & -1.0547 & 0.2916 & 0.8089\\
\bottomrule
\end{tabular}
\end{table*}

\begin{table*}[tbh]

\caption{\label{tab:kable3}OLR Summary -  Intelligence  - Ref. Levels: Type Rationale-Generation Robot and Group Non-AI Group\label{tab:Intelligence_ATurk}}
\centering
\begin{tabular}[t]{lccccc}
\toprule
  & Value & Std. Error & t value & p value & Odds Ratio\\
\midrule
Action-Declaring Robot & -0.8585 & 0.3631 & -2.3643 & 0.0181 & 0.4238\\
Numerical-Reasoning Robot & -1.0593 & 0.3716 & -2.8510 & 0.0044 & 0.3467\\
AI Group & 0.3321 & 0.3304 & 1.0053 & 0.3148 & 1.3939\\
Action-Declaring Robot:AI Group & -1.0046 & 0.4542 & -2.2119 & 0.0270 & 0.3662\\
Numerical-Reasoning Robot:AI Group & 0.1047 & 0.4626 & 0.2263 & 0.8210 & 1.1104\\
\addlinespace
1|2 & -1.4164 & 0.2748 & -5.1533 & 0.0000 & 0.2426\\
2|3 & 0.1200 & 0.2651 & 0.4528 & 0.6507 & 1.1275\\
\bottomrule
\end{tabular}
\end{table*}

\begin{table*}[tbh]

\caption{\label{tab:kable3}OLR Summary -  Intelligence  - Ref. Levels: Type Action-Declaring Robot and Group AI Group\label{tab:Intelligence_BNon-Turk}}
\centering
\begin{tabular}[t]{lccccc}
\toprule
  & Value & Std. Error & t value & p value & Odds Ratio\\
\midrule
Rationale-Generation Robot & 1.8631 & 0.2799 & 6.6555 & 0.0000 & 6.4440\\
Numerical-Reasoning Robot & 0.9085 & 0.2704 & 3.3604 & 0.0008 & 2.4806\\
Non-AI Group & 0.6725 & 0.3110 & 2.1623 & 0.0306 & 1.9592\\
Rationale-Generation Robot:Non-AI Group & -1.0047 & 0.4542 & -2.2120 & 0.0270 & 0.3662\\
Numerical-Reasoning Robot:Non-AI Group & -1.1093 & 0.4502 & -2.4641 & 0.0137 & 0.3298\\
\addlinespace
1|2 & 0.1147 & 0.1878 & 0.6106 & 0.5415 & 1.1215\\
2|3 & 1.6511 & 0.2051 & 8.0509 & 0.0000 & 5.2125\\
\bottomrule
\end{tabular}
\end{table*}

\begin{table*}[tbh]

\caption{\label{tab:kable3}OLR Summary -  Intelligence  - Ref. Levels: Type Action-Declaring Robot and Group Non-AI Group\label{tab:Intelligence_BTurk}}
\centering
\begin{tabular}[t]{lccccc}
\toprule
  & Value & Std. Error & t value & p value & Odds Ratio\\
\midrule
Rationale-Generation Robot & 0.8585 & 0.3631 & 2.3642 & 0.0181 & 2.3596\\
Numerical-Reasoning Robot & -0.2008 & 0.3593 & -0.5589 & 0.5762 & 0.8181\\
AI Group & -0.6725 & 0.3110 & -2.1623 & 0.0306 & 0.5104\\
Rationale-Generation Robot:AI Group & 1.0047 & 0.4542 & 2.2120 & 0.0270 & 2.7310\\
Numerical-Reasoning Robot:AI Group & 1.1093 & 0.4502 & 2.4641 & 0.0137 & 3.0323\\
\addlinespace
1|2 & -0.5579 & 0.2530 & -2.2047 & 0.0275 & 0.5724\\
2|3 & 0.9785 & 0.2566 & 3.8132 & 0.0001 & 2.6605\\
\bottomrule
\end{tabular}
\end{table*}

\begin{table*}[tbh]

\caption{\label{tab:kable3}OLR Summary -  Intelligence  - Ref. Levels: Type Numerical-Reasoning Robot and Group AI Group\label{tab:Intelligence_CNon-Turk}}
\centering
\begin{tabular}[t]{lccccc}
\toprule
  & Value & Std. Error & t value & p value & Odds Ratio\\
\midrule
Rationale-Generation Robot & 0.9546 & 0.2801 & 3.4088 & 0.0007 & 2.5977\\
Action-Declaring Robot & -0.9085 & 0.2704 & -3.3604 & 0.0008 & 0.4031\\
Non-AI Group & -0.4368 & 0.3244 & -1.3467 & 0.1781 & 0.6461\\
Rationale-Generation Robot:Non-AI Group & 0.1047 & 0.4626 & 0.2263 & 0.8210 & 1.1104\\
Action-Declaring Robot:Non-AI Group & 1.1093 & 0.4502 & 2.4641 & 0.0137 & 3.0323\\
\addlinespace
1|2 & -0.7939 & 0.2021 & -3.9274 & 0.0001 & 0.4521\\
2|3 & 0.7425 & 0.2016 & 3.6836 & 0.0002 & 2.1013\\
\bottomrule
\end{tabular}
\end{table*}

\begin{table*}[tbh]

\caption{\label{tab:kable3}OLR Summary -  Intelligence  - Ref. Levels: Type Numerical-Reasoning Robot and Group Non-AI Group\label{tab:Intelligence_CTurk}}
\centering
\begin{tabular}[t]{lccccc}
\toprule
  & Value & Std. Error & t value & p value & Odds Ratio\\
\midrule
Rationale-Generation Robot & 1.0593 & 0.3716 & 2.8510 & 0.0044 & 2.8844\\
Action-Declaring Robot & 0.2008 & 0.3593 & 0.5589 & 0.5763 & 1.2224\\
AI Group & 0.4368 & 0.3244 & 1.3467 & 0.1781 & 1.5477\\
Rationale-Generation Robot:AI Group & -0.1047 & 0.4626 & -0.2263 & 0.8210 & 0.9006\\
Action-Declaring Robot:AI Group & -1.1093 & 0.4502 & -2.4641 & 0.0137 & 0.3298\\
\addlinespace
1|2 & -0.3571 & 0.2628 & -1.3587 & 0.1742 & 0.6997\\
2|3 & 1.1793 & 0.2693 & 4.3787 & 0.0000 & 3.2522\\
\bottomrule
\end{tabular}
\end{table*}

\begin{table*}[tbh]

\caption{\label{tab:kable3}OLR Summary -  Potential  - Ref. Levels: Type Rationale-Generation Robot and Group AI Group\label{tab:Potential_ANon-Turk}}
\centering
\begin{tabular}[t]{lccccc}
\toprule
  & Value & Std. Error & t value & p value & Odds Ratio\\
\midrule
Action-Declaring Robot & -1.4669 & 0.2736 & -5.3619 & 0.0000 & 0.2306\\
Numerical-Reasoning Robot & -1.4049 & 0.2855 & -4.9210 & 0.0000 & 0.2454\\
Non-AI Group & 0.3139 & 0.3403 & 0.9226 & 0.3562 & 1.3688\\
Action-Declaring Robot:Non-AI Group & 0.3615 & 0.4584 & 0.7888 & 0.4303 & 1.4355\\
Numerical-Reasoning Robot:Non-AI Group & -1.2945 & 0.4854 & -2.6667 & 0.0077 & 0.2740\\
\addlinespace
1|2 & -1.7680 & 0.2168 & -8.1558 & 0.0000 & 0.1707\\
2|3 & -0.1478 & 0.1969 & -0.7505 & 0.4530 & 0.8626\\
\bottomrule
\end{tabular}
\end{table*}

\begin{table*}[tbh]

\caption{\label{tab:kable3}OLR Summary -  Potential  - Ref. Levels: Type Rationale-Generation Robot and Group Non-AI Group\label{tab:Potential_ATurk}}
\centering
\begin{tabular}[t]{lccccc}
\toprule
  & Value & Std. Error & t value & p value & Odds Ratio\\
\midrule
Action-Declaring Robot & -1.1053 & 0.3729 & -2.9642 & 0.0030 & 0.3311\\
Numerical-Reasoning Robot & -2.6994 & 0.4039 & -6.6836 & 0.0000 & 0.0672\\
AI Group & -0.3139 & 0.3403 & -0.9224 & 0.3563 & 0.7306\\
Action-Declaring Robot:AI Group & -0.3616 & 0.4584 & -0.7889 & 0.4302 & 0.6966\\
Numerical-Reasoning Robot:AI Group & 1.2944 & 0.4854 & 2.6666 & 0.0077 & 3.6489\\
\addlinespace
1|2 & -2.0820 & 0.2966 & -7.0203 & 0.0000 & 0.1247\\
2|3 & -0.4617 & 0.2792 & -1.6534 & 0.0983 & 0.6302\\
\bottomrule
\end{tabular}
\end{table*}

\begin{table*}[tbh]

\caption{\label{tab:kable3}OLR Summary -  Potential  - Ref. Levels: Type Action-Declaring Robot and Group AI Group\label{tab:Potential_BNon-Turk}}
\centering
\begin{tabular}[t]{lccccc}
\toprule
  & Value & Std. Error & t value & p value & Odds Ratio\\
\midrule
Rationale-Generation Robot & 1.4668 & 0.2736 & 5.3616 & 0.0000 & 4.3354\\
Numerical-Reasoning Robot & 0.0620 & 0.2728 & 0.2271 & 0.8203 & 1.0639\\
Non-AI Group & 0.6755 & 0.3077 & 2.1950 & 0.0282 & 1.9649\\
Rationale-Generation Robot:Non-AI Group & -0.3614 & 0.4584 & -0.7885 & 0.4304 & 0.6967\\
Numerical-Reasoning Robot:Non-AI Group & -1.6561 & 0.4643 & -3.5668 & 0.0004 & 0.1909\\
\addlinespace
1|2 & -0.3012 & 0.1880 & -1.6022 & 0.1091 & 0.7399\\
2|3 & 1.3191 & 0.2005 & 6.5806 & 0.0000 & 3.7401\\
\bottomrule
\end{tabular}
\end{table*}

\begin{table*}[tbh]

\caption{\label{tab:kable3}OLR Summary -  Potential  - Ref. Levels: Type Action-Declaring Robot and Group Non-AI Group\label{tab:Potential_BTurk}}
\centering
\begin{tabular}[t]{lccccc}
\toprule
  & Value & Std. Error & t value & p value & Odds Ratio\\
\midrule
Rationale-Generation Robot & 1.1053 & 0.3729 & 2.9641 & 0.0030 & 3.0201\\
Numerical-Reasoning Robot & -1.5941 & 0.3754 & -4.2465 & 0.0000 & 0.2031\\
AI Group & -0.6755 & 0.3077 & -2.1951 & 0.0282 & 0.5089\\
Rationale-Generation Robot:AI Group & 0.3616 & 0.4584 & 0.7889 & 0.4302 & 1.4356\\
Numerical-Reasoning Robot:AI Group & 1.6560 & 0.4643 & 3.5667 & 0.0004 & 5.2385\\
\addlinespace
1|2 & -0.9767 & 0.2543 & -3.8407 & 0.0001 & 0.3766\\
2|3 & 0.6436 & 0.2510 & 2.5639 & 0.0104 & 1.9034\\
\bottomrule
\end{tabular}
\end{table*}

\begin{table*}[tbh]

\caption{\label{tab:kable3}OLR Summary -  Potential  - Ref. Levels: Type Numerical-Reasoning Robot and Group AI Group\label{tab:Potential_CNon-Turk}}
\centering
\begin{tabular}[t]{lccccc}
\toprule
  & Value & Std. Error & t value & p value & Odds Ratio\\
\midrule
Rationale-Generation Robot & 1.4049 & 0.2855 & 4.9210 & 0.0000 & 4.0752\\
Action-Declaring Robot & -0.0620 & 0.2728 & -0.2271 & 0.8203 & 0.9399\\
Non-AI Group & -0.9806 & 0.3456 & -2.8371 & 0.0046 & 0.3751\\
Rationale-Generation Robot:Non-AI Group & 1.2945 & 0.4854 & 2.6667 & 0.0077 & 3.6491\\
Action-Declaring Robot:Non-AI Group & 1.6560 & 0.4643 & 3.5667 & 0.0004 & 5.2386\\
\addlinespace
1|2 & -0.3631 & 0.2060 & -1.7628 & 0.0779 & 0.6955\\
2|3 & 1.2571 & 0.2162 & 5.8156 & 0.0000 & 3.5154\\
\bottomrule
\end{tabular}
\end{table*}

\begin{table*}[tbh]

\caption{\label{tab:kable3}OLR Summary -  Potential  - Ref. Levels: Type Numerical-Reasoning Robot and Group Non-AI Group\label{tab:Potential_CTurk}}
\centering
\begin{tabular}[t]{lccccc}
\toprule
  & Value & Std. Error & t value & p value & Odds Ratio\\
\midrule
Rationale-Generation Robot & 2.6994 & 0.4039 & 6.6837 & 0.0000 & 14.8709\\
Action-Declaring Robot & 1.5941 & 0.3754 & 4.2466 & 0.0000 & 4.9238\\
AI Group & 0.9806 & 0.3456 & 2.8371 & 0.0046 & 2.6659\\
Rationale-Generation Robot:AI Group & -1.2945 & 0.4854 & -2.6667 & 0.0077 & 0.2740\\
Action-Declaring Robot:AI Group & -1.6560 & 0.4643 & -3.5667 & 0.0004 & 0.1909\\
\addlinespace
1|2 & 0.6174 & 0.2798 & 2.2063 & 0.0274 & 1.8541\\
2|3 & 2.2377 & 0.2983 & 7.5020 & 0.0000 & 9.3718\\
\bottomrule
\end{tabular}
\end{table*}

\begin{table*}[tbh]

\caption{\label{tab:kable3}OLR Summary -  Understandability  - Ref. Levels: Type Rationale-Generation Robot and Group AI Group\label{tab:Understandability_ANon-Turk}}
\centering
\begin{tabular}[t]{lccccc}
\toprule
  & Value & Std. Error & t value & p value & Odds Ratio\\
\midrule
Action-Declaring Robot & -2.6153 & 0.3274 & -7.9873 & 0.0000 & 0.0731\\
Numerical-Reasoning Robot & -4.2054 & 0.3661 & -11.4866 & 0.0000 & 0.0149\\
Non-AI Group & -0.4175 & 0.3664 & -1.1393 & 0.2546 & 0.6587\\
Action-Declaring Robot:Non-AI Group & 1.5455 & 0.4949 & 3.1227 & 0.0018 & 4.6903\\
Numerical-Reasoning Robot:Non-AI Group & -1.7585 & 0.7309 & -2.4059 & 0.0161 & 0.1723\\
\addlinespace
1|2 & -3.5627 & 0.3026 & -11.7746 & 0.0000 & 0.0284\\
2|3 & -1.0644 & 0.2358 & -4.5135 & 0.0000 & 0.3449\\
\bottomrule
\end{tabular}
\end{table*}

\begin{table*}[tbh]

\caption{\label{tab:kable3}OLR Summary -  Understandability  - Ref. Levels: Type Rationale-Generation Robot and Group Non-AI Group\label{tab:Understandability_ATurk}}
\centering
\begin{tabular}[t]{lccccc}
\toprule
  & Value & Std. Error & t value & p value & Odds Ratio\\
\midrule
Action-Declaring Robot & -1.0698 & 0.3807 & -2.8100 & 0.0050 & 0.3431\\
Numerical-Reasoning Robot & -5.9638 & 0.6824 & -8.7391 & 0.0000 & 0.0026\\
AI Group & 0.4175 & 0.3664 & 1.1393 & 0.2546 & 1.5182\\
Action-Declaring Robot:AI Group & -1.5455 & 0.4949 & -3.1227 & 0.0018 & 0.2132\\
Numerical-Reasoning Robot:AI Group & 1.7584 & 0.7309 & 2.4058 & 0.0161 & 5.8032\\
\addlinespace
1|2 & -3.1452 & 0.3361 & -9.3583 & 0.0000 & 0.0431\\
2|3 & -0.6469 & 0.2808 & -2.3041 & 0.0212 & 0.5237\\
\bottomrule
\end{tabular}
\end{table*}

\begin{table*}[tbh]

\caption{\label{tab:kable3}OLR Summary -  Understandability  - Ref. Levels: Type Action-Declaring Robot and Group AI Group\label{tab:Understandability_BNon-Turk}}
\centering
\begin{tabular}[t]{lccccc}
\toprule
  & Value & Std. Error & t value & p value & Odds Ratio\\
\midrule
Rationale-Generation Robot & 2.6151 & 0.3274 & 7.9869 & 0.0000 & 13.6684\\
Numerical-Reasoning Robot & -1.5903 & 0.2993 & -5.3142 & 0.0000 & 0.2039\\
Non-AI Group & 1.1277 & 0.3316 & 3.4009 & 0.0007 & 3.0884\\
Rationale-Generation Robot:Non-AI Group & -1.5454 & 0.4949 & -3.1226 & 0.0018 & 0.2132\\
Numerical-Reasoning Robot:Non-AI Group & -3.3039 & 0.7154 & -4.6183 & 0.0000 & 0.0367\\
\addlinespace
1|2 & -0.9474 & 0.2107 & -4.4964 & 0.0000 & 0.3877\\
2|3 & 1.5506 & 0.2308 & 6.7170 & 0.0000 & 4.7143\\
\bottomrule
\end{tabular}
\end{table*}

\begin{table*}[tbh]

\caption{\label{tab:kable3}OLR Summary -  Understandability  - Ref. Levels: Type Action-Declaring Robot and Group Non-AI Group\label{tab:Understandability_BTurk}}
\centering
\begin{tabular}[t]{lccccc}
\toprule
  & Value & Std. Error & t value & p value & Odds Ratio\\
\midrule
Rationale-Generation Robot & 1.0698 & 0.3807 & 2.8099 & 0.0050 & 2.9147\\
Numerical-Reasoning Robot & -4.8941 & 0.6646 & -7.3645 & 0.0000 & 0.0075\\
AI Group & -1.1281 & 0.3316 & -3.4020 & 0.0007 & 0.3237\\
Rationale-Generation Robot:AI Group & 1.5456 & 0.4949 & 3.1229 & 0.0018 & 4.6909\\
Numerical-Reasoning Robot:AI Group & 3.3039 & 0.7153 & 4.6190 & 0.0000 & 27.2187\\
\addlinespace
1|2 & -2.0755 & 0.2980 & -6.9645 & 0.0000 & 0.1255\\
2|3 & 0.4228 & 0.2591 & 1.6321 & 0.1027 & 1.5263\\
\bottomrule
\end{tabular}
\end{table*}

\begin{table*}[tbh]

\caption{\label{tab:kable3}OLR Summary -  Understandability  - Ref. Levels: Type Numerical-Reasoning Robot and Group AI Group\label{tab:Understandability_CNon-Turk}}
\centering
\begin{tabular}[t]{lccccc}
\toprule
  & Value & Std. Error & t value & p value & Odds Ratio\\
\midrule
Rationale-Generation Robot & 4.2054 & 0.3661 & 11.4867 & 0.0000 & 67.0472\\
Action-Declaring Robot & 1.5901 & 0.2993 & 5.3136 & 0.0000 & 4.9043\\
Non-AI Group & -2.1759 & 0.6325 & -3.4403 & 0.0006 & 0.1135\\
Rationale-Generation Robot:Non-AI Group & 1.7584 & 0.7309 & 2.4058 & 0.0161 & 5.8031\\
Action-Declaring Robot:Non-AI Group & 3.3039 & 0.7153 & 4.6189 & 0.0000 & 27.2188\\
\addlinespace
1|2 & 0.6427 & 0.2166 & 2.9667 & 0.0030 & 1.9015\\
2|3 & 3.1410 & 0.2846 & 11.0345 & 0.0000 & 23.1261\\
\bottomrule
\end{tabular}
\end{table*}

\begin{table*}[tbh]

\caption{\label{tab:kable3}OLR Summary -  Understandability  - Ref. Levels: Type Numerical-Reasoning Robot and Group Non-AI Group\label{tab:Understandability_CTurk}}
\centering
\begin{tabular}[t]{lccccc}
\toprule
  & Value & Std. Error & t value & p value & Odds Ratio\\
\midrule
Rationale-Generation Robot & 5.9638 & 0.6824 & 8.7390 & 0.0000 & 389.0826\\
Action-Declaring Robot & 4.8940 & 0.6646 & 7.3643 & 0.0000 & 133.4890\\
AI Group & 2.1759 & 0.6325 & 3.4403 & 0.0006 & 8.8105\\
Rationale-Generation Robot:AI Group & -1.7584 & 0.7309 & -2.4058 & 0.0161 & 0.1723\\
Action-Declaring Robot:AI Group & -3.3039 & 0.7153 & -4.6189 & 0.0000 & 0.0367\\
\addlinespace
1|2 & 2.8186 & 0.5943 & 4.7431 & 0.0000 & 16.7534\\
2|3 & 5.3169 & 0.6257 & 8.4981 & 0.0000 & 203.7525\\
\bottomrule
\end{tabular}
\end{table*}
\end{document}